\documentclass[showpacs,superscriptaddress,groupedaddress,showkeys]{revtex4-1}
\usepackage[top=2.5cm, bottom=2.5cm, left=2.5cm, right=2.5cm,headheight=15pt]{geometry} 

\usepackage{graphicx}
\usepackage{bm}%
\usepackage{amsmath} 
\usepackage[caption=false]{subfig}
\usepackage{hyperref}
\setcitestyle{authoryear,round}
\setcitestyle{citesep={;}}

\begin{document}
\title{Extension of a Reduced Entropic Model of Electron Transport to Magnetized Nonlocal Regimes of High-Energy-Density Plasmas}
\author{D. Del Sorbo}
\author{J.-L. Feugeas}
\author{Ph. Nicola\"i}
\affiliation{Universit\'e de Bordeaux-CNRS-CEA, Centre Lasers Intenses et Applications, UMR 5107, F-33405 Talence, France}
\author{M. Olazabal-Loum\'e}
\affiliation{CEA/CESTA, F-33114 Le Barp, France }
\author{B. Dubroca}
\author{V. Tikhonchuk}
\affiliation{Universit\'e de Bordeaux-CNRS-CEA, Centre Lasers Intenses et Applications, UMR 5107, F-33405 Talence, France}
\keywords{Inertial Confinement Fusion; High-Energy-Density Physics; Nonlocal Electron Transport; Magnetized Plasmas; Kinetic Model}

\clearpage
    
\begin{abstract}

Laser produced high-energy-density plasmas may contain strong magnetic fields that affect the energy transport, which can be nonlocal. Models which describe the magnetized nonlocal transport are formally complicated and based on many approximations. This paper presents a more straightforward approach to the description of the electron transport in this regime, based on the extension of a reduced entropic model. The calculated magnetized heat fluxes are compared with the known asymptotic limits and applied for studying of a magnetized nonocal plasma thermalization. 
    \end{abstract}

    \maketitle

\section{Introduction}

Magnetic fields play an important role in astrophysics \citep{gregori2012generation,marocchino2015magnetic} and may affect the transport properties of high-energy-density plasmas in laboratory conditions \citep{stamper1971spontaneous,clark2000measurements}.

In laser-produced plasmas, magnetic fields can be self-generated \citep{kingham2002nonlocal} or externally imposed \citep{froula2007quenching}. In the inertial confinement fusion context, external magnetic fields may limit energy losses from the central hot spot or guide the energy flux to the central part of the target \citep{strozzi2012fast,cuneo2012magnetically}. External fields could also modify the target symmetry during implosion, as well as the energy transfer through heat conduction zone. The electron conduction model implemented in the majority of radiation hydrodynamic codes is based on the assumption of a small deviation from the local thermodynamic equilibrium \citep{spitzer,Braginskii1965}. But, if the electron mean free path (MFP) exceeds a fraction of  the temperature gradient length, this assumption is no more valid and a nonlocal theory is required \citep{malone}.
This is a common case in the inertial confinement fusion context.

Kinetic codes \citep{bell2006fast,tzoufras2011vlasov} are able to describe nonlocal transport, even in magnetized plasmas. However, they are too computationally expensive to provide results in the time scale of interest for hydrodynamics, either by a direct computation or by coupling them with hydrodynamic codes. For this reason, simplified descriptions, which capture the main features of nonlocal transport, have been derived \citep{bell1985non,manheimer2008development}. We refer to them as nonlocal models.

The description of magnetic field effects on the electron heat transport is complicated \citep{Braginskii1965,epperlein1986plasma}. It involves tensor transport coefficients, which account for flux limitation effects and flux rotation.
 One of the most frequently used nonlocal electron transport model  \citep{meezan2004hydrodynamics,hu2008validation,craxton2015direct,marocchino},  proposed by Schurtz, Nicola\"i and Busquet \citep{guy} (SNB), is based on a simplified kinetic approach: the nonlocal electron contribution is computed  using a diffusion equation, resolved with a multigroup scheme on energy. It has been generalized to the magnetized nonlocal regime, thanks to a phenomenological modeling of the induced electric field \citep{nicolai}. Nevertheless, despite this simplification, the model remains too complex and not reliable to be used in practice.

 In this paper we propose  to approach the problem in a different way, by using a first-principles model,  already tested in the nonlocal  regime \citep{del2015reduced}, without magnetic fields. This model, so-called M1, is based on a simplified Fokker-Planck (FP) equation, developed on angular moments and closed with the local angular entropy maximization principle,  which is different from the standard spherical harmonic development.  Such a closure has already been tested for strongly anisotropic beams of particles \citep{touati} in domains of fast ignition \citep{del2015approach} and radiotherapy \citep{jejeegaga}.  The M1 model presents the best properties to be generalized for the study of the magnetized nonlocal transport due to its simple and numerically stable mathematical structure.   

The M1 predictions for a magnetized electron energy transport are studied in a broad range of parameters covering the local as well as nonlocal regime. It is shown that the model results agree with known asymptotic expressions. The model is applied to studying the evolution of a thermal wave in the magnetized nonlocal regime.

The paper is organized as follows. In Sec.~\ref{Local electron transport theories}, an overview of unmagnetized and magnetized local transport models is presented; in Sec.~\ref{The reduced entropic model}, we derive the entropic model using the induced electric field response developed in Sec.~\ref{Nonlocal electromagnetic fields}. The model is applied in Sec.~\ref{Local regime of magnetized plasmas}, to recover the local limit and compare it with known theoretical results. The nonlocal limit is presented in Sec.~\ref{Magnetized nonlocal heat transport}. In Sec.~\ref{Relaxation to thermal equilibrium} the model is applied for studying of evolution of magnetized thermal waves. The conclusions are presented in Sec.~\ref{On the magnetized transport}.

\section{Local electron transport theories}\label{Local electron transport theories}

The electron heat flux can be calculated knowing the electron distribution function (EDF):
\begin{equation}
\vec{q}=\int_{\mathbf{R^{3}}}d^{3}v\frac{m_{e}v^{2}}{2}\vec{v}{f}_{e}.\label{heat flux definition}
\end{equation}
 The EDF temporal evolution is given by the Landau-FP equation \citep{landau10}
\begin{equation}
\left(\frac{\partial}{\partial t}+\vec{v}\cdot \vec{\nabla}+\vec{a}\cdot \vec{\nabla}_{v}\right)f_{e}=\left(\frac{\partial f_{e}}{\partial t}\right)_{\mbox{coll}},\label{fokker plank}
\end{equation}
where $t$ is the time, $\vec{v}$ is the velocity, $\vec{a}=e/m_{e}(\vec{E}+\vec{v}/c\times\vec{B})$ is the Lorentz acceleration, function of the electric $\vec{E}$ and magnetic $\vec{B}$ fields. The elementary charge is $e$, the electron mass is $m_{e}$, the light speed is $c$ and $\left({\partial f_{e}}/{\partial t}\right)_{\mbox{coll}}$ is a term accounting for collision modeling.

The classical theory of electron transport is based on the assumption of a small deviation from the local thermodynamic equilibrium,  that is, the EDF is composed of the Maxwellian distribution $f_{e}^{m}=n_{e}/(2\pi v_{th}^{2})^{3/2}e^{-v^{2}/(2v_{th}^{2})}$ and a linear anisotropic perturbation, in the velocity direction $\vec{\Omega}=\vec{v}/v$. Here, $n_{e}$ is the electron density, $T_{e}$ the electron temperature and $v_{th}=\sqrt{T_{e}/m_{e}}$ is the thermal velocity. In this approximation, the EDF reads
\begin{equation}
f_{e}=f_{e}^{m}+\frac{3}{4\pi}\vec{\Omega}\cdot\vec{f}_{1},\label{local approximation}
\end{equation}
where $\vec{f}_{1}$ is the first angular moment of the EDF, such that $||\vec{f}_{1}||\ll f_{e}^{m}$.

The Maxwellian hypothesis for the main part of the EDF, yields to the so-called local transport theory.

A local theory of the magnetized electron transport has been derived by Braginskii \citep{Braginskii1965} and further corrected by Epperlein and Haines \citep{epperlein1986plasma}.
Approximate solutions of the local FP equation allowed to calculate thermoelectric coefficients for all magnetohydrodynamic processes. These coefficients provide a relation between the electric field, the temperature gradient, the current and the electron heat flux. They read respectively \citep{Braginskii1965,epperlein1986plasma,nicolai}:
\begin{equation}
\vec{E}=
-\frac{\vec{\nabla}p_{e}}{en_{e}}
+\frac{\vec{j}_{e}\times{\vec{B}}}{cen_{e}}
+\bar{\bar{\alpha}}\cdot\frac{\vec{j}_{e}}{{e}^{2}n_{e}^{2}}
-\bar{\bar{\beta}}\cdot\frac{\vec{\nabla}T_{e}}{e}\label{E braginskii}
\end{equation}
and
\begin{equation}
\vec{q}_{B}=
-\bar{\bar{k}}\cdot\vec{\nabla}T_{e}
-\bar{\bar{\beta}}\cdot\vec{j}_{e}\frac{T_{e}}{e},\label{q braginskii}
\end{equation}
where the current is provided by the stationary Ampere's law
\begin{equation}
\vec{j}_{e}=\frac{c}{4\pi}\vec{\nabla}\times\vec{B}.\label{ampere's law}
\end{equation}

In Eqs.~\eqref{E braginskii} and \eqref{q braginskii}, $\bar{\bar{\alpha}}$, $\bar{\bar{\beta}}$ and $\bar{\bar{k}}$ are second-order tensors, which respectively account for electrical resistivity, thermoelectric and thermal conductivity.
Applied to a generic vector $\vec{V}$, this tensor quantities (generically denoted by $\bar{\bar{\Psi}}$) can be expressed in terms of components relative to the magnetic field unitary vector $\hat{b}=\vec{B}/B$, as
\begin{equation}
\bar{\bar{\Psi}}\cdot\vec{V}=
\Psi_{\parallel}\hat{b}\cdot(\hat{b}\cdot\vec{V})
-\Psi_{\perp}\hat{b}\times(\hat{b}\times\vec{V})
\pm\Psi_{\wedge}\hat{b}\times\vec{V},
\end{equation}
where the negative sign applies only to the case $\bar{\bar{\Psi}}=\bar{\bar{\alpha}}$. 
The symbols $\parallel$, $\perp$ mean parallel and perpendicular to the magnetic field, while $\wedge$ means the direction perpendicular to the magnetic field and the generic vector.

The Braginskii's transport theory is a practical way to account for magnetic effects in hydrodynamic codes. The numerical value for the transport coefficients has been later improved by Epperlein and Haines \citep{epperlein1986plasma}. They have reduced the error on  Braginskii's coefficients up to $15\%$, deducing them numerically.

The role of magnetic fields in electron transport is characterized by a magnetization parameter (the Hall parameter), which is a product of the electron collision time $\tau_{e}=\nu_{e}^{-1}=3{\sqrt{m_{e}T_{e}^{3}}}/({4\sqrt{2\pi}n_{e}Ze^{4}\Lambda_{ei}})$, where $\Lambda_{ei}$ is the Coulomb logarithm, and the electron gyrofrequency $\omega_{B}={eB}/({m_{e}c})$.

In the limit $\omega_{B}\tau_{e}= 0$, the Braginskii's theory reduces to the Spitzer and H\"arm (SH) theory \citep{spitzer}. The Ampere's law \eqref{ampere's law} is replaced by the zero current condition
$\vec{j}_{e}=-e\int_{\mathbf{R}^{3}}d^{3}v\vec{v}f_{e}=0$, so the electric field \eqref{E braginskii} becomes
\begin{equation}
\vec{E}_{SH}=-\frac{T_{e}}{e}\left(\frac{\vec{\nabla}n_{e}}{n_{e}}+\xi\frac{\vec{\nabla}T_{e}}{T_{e}}\right),\label{equilibrium E}
\end{equation}
with $\xi(Z)=1+\beta=1+{3}/{2}(Z+0.477)/(Z+2.15)$.
For simplicity we consider electrons as a perfect gas with $p_{e}=n_{e}T_{e}$. Then, the electron heat flux \eqref{q braginskii} contains only one term, with the thermal conductivity.

\section{Reduced entropic model}\label{The reduced entropic model}

The purpose of this paper is to describe the combined effect of nonlocal transport and of magnetic fields on the heat flux. 

The SH theory predicts that the heat is transported by suprathermal electrons having the velocity $\sim3.7$ times higher than the thermal velocity \citep{guy}.
Since the MFP varies as $\propto v^{4}$, the heat is transported by electrons having the collision length $187$ times larger than the thermal MFP.
These electrons can penetrate deeply into the plasma and can deposit their energy far away from where they have been originated.
In this case, the local thermodynamic equilibrium cannot be reached.
In particular, if the MFP is long compared to the temperature gradient length ($\lambda_{e}>2\times 10^{-3}L_{T}$), the local theory is no more valid and the heat transport becomes nonlocal, requiring a kinetic treatment \citep{bell2006fast,tzoufras2011vlasov}.

Using kinetic codes, the complexity of combining nonlocal and magnetic effects mainly limits the analysis to kinetic temporal scales (few tenths of collision times) and to a reduced number of dimensions. The entropic model M1 \citep{del2015reduced}, based on a reduced FP equation has been modified in such a way that it can be coupled with magnetohydrodynamic codes, thus extending the analysis to long hydrodynamic times and to multidimensional geometries. In particular, our purpose is to test the validity of this improved version of the entropic model, in the magnetized nonlocal regime.


Since the electrons transporting the heat are suprathermal, following Refs.~\citep{del2015reduced,albritton}, we limit our analysis to fast electrons colliding with thermal ions and electrons. The collision integral is taken in a simplified form, proposed by Albritton et al.~\citep{albritton,albritton2} and applied in Ref.~\citep{del2015reduced} to unmagnetized plasmas. 
Moreover, following to Refs.~\citep{touati, dubroca}, we developed the EDF in angular moments and retain the first two moments in the M1 approximation. As the electron collision time  is short compared to the characteristic hydrodynamic time, we solve a stationary kinetic equation at each hydrodynamic time step. Then the system of kinetic equations for two angular moments of the EDF writes:
\begin{widetext}
\begin{equation}
\begin{cases}
v\vec{\nabla}\cdot\vec{f}_{1}-\frac{e\vec{E}}{m_{e}v^{2}} \cdot\frac{\partial}{\partial v}\left(v^{2}\vec{f}_{1}\right)=\nu_{ee}v\frac{\partial}{\partial v}\left(f_{0}-f_{0}^{m}\right)
\\
v \vec{\nabla}\cdot\bar{\bar{f}}_{2}-\frac{e}{m_{e}v^{2}}\frac{\partial}{\partial v}\left(v^{2}\bar{\bar{f}}_{2}\cdot\vec{E}\right)+\frac{e}{m_{e}v}\left(f_{0}\bar{\bar{I}}-\bar{\bar{f}}_{2}\right)\cdot\vec{E}+\frac{e}{m_{e}c}\vec{f}_{1}\times \vec{B}=\nu_{ee}v\frac{\partial}{\partial v}\vec{f}_{1}
-
(\nu_{ee}+\nu_{ei})\vec{f}_{1}
\\
\bar{\bar{f}}_{2}=\left[\frac{1}{3}\bar{\bar{I}}+\frac{\vec{\Omega}_{v}^{2}}{2}\left(1+\vec{\Omega}_{v}^{2}\right)\left(\frac{\vec{f}_{1}\otimes\vec{f}_{1}}{\vec{f}_{1}^{2}}-\frac{1}{3}\bar{\bar{I}}\right)\right]f_{0}
\end{cases}.\label{first two moments}
\end{equation}
\end{widetext}
Third equation presents a closure relation of the second angular moment. It is derived by maximizing the local angular entropy
\begin{equation}
H_{\Omega}(\vec{x},v)=-\int_{S^{2}}d^{2}\Omega\left(f_{e}\log f_{e}-f_{e}\right)\; \forall (\vec{x},v),\label{local angular entropy}
\end{equation} 
imposing the dependence on the first two moments. 
Here $\bar{\bar{I}}$ is the second-order identity tensor, 
and $\vec{\Omega}_{v}=\vec{f}_{1}/f_{0}$.

 Starting from the kinetic point of view, the hydrodynamics can be constructed from the maximization of the local entropy $\int_{\mathbf{R}^{+}}dvv^{2}H_{\Omega}(\vec{x},v)\; \forall \vec{x}$. The maximization of the local angular entropy \eqref{local angular entropy} is less restrictive, it presents  a compromise which allows to account for kinetic effects in a mesoscopic way.

 The M1 model can be coupled to a multidimensional magnetohydrodynamic code as follows: at every time step of a magnetohydrodynamic simulation it is initialized by the hydrodynamic parameters (temperature, density and magnetic field) which are used to compute the EDF. The latter can be easily integrated to obtain the heat flux. We stress that the M1 model does not perform any distinction between local and nonlocal populations but it computes the EDF of the plasma as a whole with the constraint of the hydrodynamic temperature. For more details, see also Sec.~IV of Ref.~\citep{del2015reduced}.

 Only one other attempt to describe the nonlocal magnetized regime on the multidimensional hydrodynamic scale has been published: the generalization of the SNB model to magnetized plasmas \citep{nicolai}. This model, however, presents many limitations. In fact the SNB model resolves a diffusion equation, which, in the magnetized regime, is complicated due to the spatial anisotropy and the appearance of many new terms. These terms are treated phenomenologically, still remaining very complex: the energy-diffusion matrix for the numerical solution is non-symmetric and so difficult to invert. On the contrary, the M1 model, which is based on a system of convective equations, remains as simple as in the unmagnetized case, described in Ref.~\citep{del2015reduced}, presenting an easier mathematical form, which is numerically solved explicitly on energy.

In addition to the temperature gradient, magnetic fields add a new direction of anisotropy to the thermal transport. Being a first moment model, the M1 model is limited to the description of one direction of anisotropy, hence it is valid if thermoelectric effects are negligible. Fortunately, this occurs frequently in laser-matter interactions.

\section{Electric and magnetic field calculations}\label{Nonlocal electromagnetic fields}

The EDF strongly depends on the electric and magnetic fields. 
The electric field $\vec{E}$ needs to be evaluated self-consistently while solving the kinetic Eq.~\eqref{first two moments}.

 The inputs of the M1 model are the hydrodynamic variables. Once they are acquired (from the previous temporal step of an hydrodynamic simulation or externally imposed), the electric field can be computed by iterating the solution of the system \eqref{first two moments},  starting from the zero or local value.
The iteration process can be summarized as follows:
 \begin{itemize}
 \item $\vec{E}$ is evaluated from the previous iteration step;
  \item Equations \eqref{first two moments} are solved for $f_{0}$ and $\vec{f}_{1}$ by using the evaluated $\vec{E}$ field;
        \item $f_{0}$ and $\vec{f}_{1}$ are used to calculate macroscopic quantities, such as $\vec{q}$ and $\vec{j}_{e}$, from which we deduce $\vec{E}$. 
\end{itemize}
The process is stopped when the electric field converges.

 The reader should be careful to do not confuse the time step of an hydrodynamic simulation from the iteration step in the resolution of system \eqref{first two moments}. Note also that the latter is formally stationary. 

 The magnetic field  $\vec{B}$ is considered in the system \eqref{first two moments} as an external source, which is assumed constant at the kinetic temporal scale ($\sim \tau_{e}$).

\subsection{Unmagnetized plasmas}

In case of unmagnetized plasmas, without external sources, Eq.~\eqref{ampere's law} reduces to $\vec{j}_{e}=0$. 
This condition implies that there are two opposite electron fluxes: suprathermal electrons which transport the heat and a slower return current flux. Since the second flux involves lower velocities, it is  less susceptible to nonlocal effects \citep{nicolai}. Thus, for weakly nonlocal conditions, the electric fields could be determined from the local Eq.~\eqref{equilibrium E}.

In the case of strong temperature gradients, the model needs to account for nonlocal corrections to the electric field. Under the assumption of a Lorentz gas  and  of a weak anisotropy (P1 approximation), the second equation of the system \eqref{first two moments} reads
\begin{equation}
\frac{v}{3} \vec{\nabla}f_{0}-\frac{e\vec{E}}{3m_{e}}\frac{\partial}{\partial v}f_{0}=-\nu_{ei}\vec{f}_{1}.
\end{equation}
A nonlocal electric field can be derived by applying the zero current condition. It has been extended to low-Z plasmas \citep{del2015reduced}:
\begin{equation}
\vec{E}_{NL}=-\frac{\xi}{2.5}\frac{m_{e}}{6e}\frac{\int_{0}^{\infty}\vec{\nabla}f_{0}v^{7}dv}{\int_{0}^{\infty}f_{0}v^{5}dv}.\label{prima definizione campo elettrico nonlocale}
\end{equation}
As shown in Ref.~\citep{del2015reduced}, for low-Z plasmas, the last expression is valid in the limit $\vec{\nabla}T_{e}/T_{e}\gg\vec{\nabla}n_{e}/n_{e}$. This condition is often respected in practice. In the extreme case of $Z=1$ and $\vec{\nabla}T_{e}/T_{e}\ll\vec{\nabla}n_{e}/n_{e}$, the error is $\sim30\%$, smaller than the one committed by models which assume local electric fields.

\subsection{Magnetized plasmas}

In magnetized plasmas, the zero-current condition is replaced by the stationary Ampere's law \eqref{ampere's law}. Unfortunately, it is not anymore possible to find an analytic expression for the electric field from Eqs.~\eqref{first two moments}. 

Magnetic fields tend to localize the transport \citep{nicolai,brantov2003linear}. Following the example of unmagnetized plasma, one may calculate electric fields in the local approximation, given by Eq.~\eqref{E braginskii}. However, it is not sufficient in the nonlocal regime. We improve this relation by using the plasma parameters updated from solution of the kinetic equation: the electron density $n_{K}=\int_{0}^{\infty}dvv^{2}f_{0}$ and the electron pressure $p_{K}={2}/{3}\int_{0}^{\infty}dvv^{2}\epsilon f_{0}$. These quantities differ from classical hydrodynamic ones because they are computed by averaging the EDF and not by solving the hydrodynamic equations.
According to the perfect gas equation of state, the kinetic electron temperature reads $T_{K}=p_{K}/n_{K}$. Using these definitions in  Eq.~\eqref{E braginskii}, the expression for the local-kinetic electric field reads
\begin{equation}
\vec{E}_{Lk}=
-\frac{\vec{\nabla}p_{K}}{en_{K}}
+\frac{\vec{j}_{e}\times{\vec{B}}}{cen_{K}}
+\bar{\bar{\alpha}}_{K}\cdot\frac{\vec{j}_{e}}{{e}^{2}n_{K}^{2}}
-\bar{\bar{\beta}}_{K}\cdot\frac{\vec{\nabla}T_{K}}{e}.\label{magnetized electric field model}
\end{equation}
The kinetic electrical resistivity is defined in function of the kinetic-hydrodynamic quantities\citep{phddario}: $\bar{\bar{\alpha}}_{K}=\bar{\bar{\alpha}}(n_{K}, T_{K})=m_{e}n_{K}/\tau_{e}(T_{K},n_{K})\bar{\bar{\alpha}}_{K}^{c}$, where $\bar{\bar{\alpha}}_{K}^{c}$ is the dimensionless electric resistivity \citep{epperlein1986plasma}.

While solving the kinetic system \eqref{first two moments}, the electric field as a function of $f_{0}$ and $\vec{f}_{1}$ is computed by iterations, according to Eq.~\eqref{magnetized electric field model}, till convergence.  Note that this new formulation does not add any degree of complexity with respect to Eq.~\eqref{prima definizione campo elettrico nonlocale}: both are integrals of the EDF and both can be computed by iterating the solution of the system \eqref{first two moments} till the field convergence.

 The electric field \eqref{magnetized electric field model} is the one that will be used along the paper.

For the treatment of magnetic fields, we consider two time scales: the hydrodynamic scale, much longer than the kinetic one. At the kinetic scale, magnetic fields are considered constant, while they vary at the hydrodynamic temporal scale according to the magnetohydrodynamic equations. 
 In the M1 model, they can be externally imposed or provided from the magnetohydrodynamic simulation to which it should be coupled. This treatment is the same used in the SNB model \citep{nicolai} and in the Braginskii's theory\citep{Braginskii1965}. Note that the addition of these fields does not require any modification to the numerical structure developed in Ref.~\citep{del2015approach}.

\section{Local regime of magnetized plasmas}\label{Local regime of magnetized plasmas}

 The M1 model, as well as the Braginskii's theory, has been presented in a multidimensional formalism. However, in what follows the analysis will be limited to one and two dimensions because our aim is to couple transport models to the CHIC code \citep{CHIC}, a two dimensional hydrodynamic code. Nevertheless, no modifications are required in order to generalize the M1 model to three dimensions.

Let us first study the local regime of magnetized plasmas. We consider a fully-ionized plasma, with a constant density, characterized by a steep temperature gradient 
\begin{equation}
T_{e}(x)=\frac{|T_{0}-T_{1}|}{2}  \left[\frac{2}{\pi}\arctan\left(\frac{x}{\delta_{NL}}\right)+1\right]+T_{1},\label{temperature atg}
\end{equation}
 with $T_{0}=5 \, \rm keV$, $T_{1}=0.5 \, \rm keV$ and $\delta_{NL}=500 \,  \mu \rm{m}$.
The system is simulated in one dimension (x-direction) with a perpendicular magnetic field (z-direction). For reasons of comparison with theory, we consider the Hall parameter $\omega_{B}\tau_{e}$ constant along the plasma, thus the magnetic field varies along with the electron collision time:
\begin{equation}
B_{z}=\frac{m_{e}c}{e\tau_{e}}\omega_{B}\tau_{e}.\label{campo magnetico test}
\end{equation}

    We consider two perfect gas plasmas, one with $Z=1$  and $n_{e}=10^{23}\;\rm cm^{-3}$ and  the second with $Z\gg1$ ($Z=79$) and $n_{e}=4\times10^{22}\;\rm cm^{-3}$.

\begin{figure}
    \centering
\subfloat[\label{kperp}]{
        \centering
        \includegraphics[width=0.45\columnwidth]{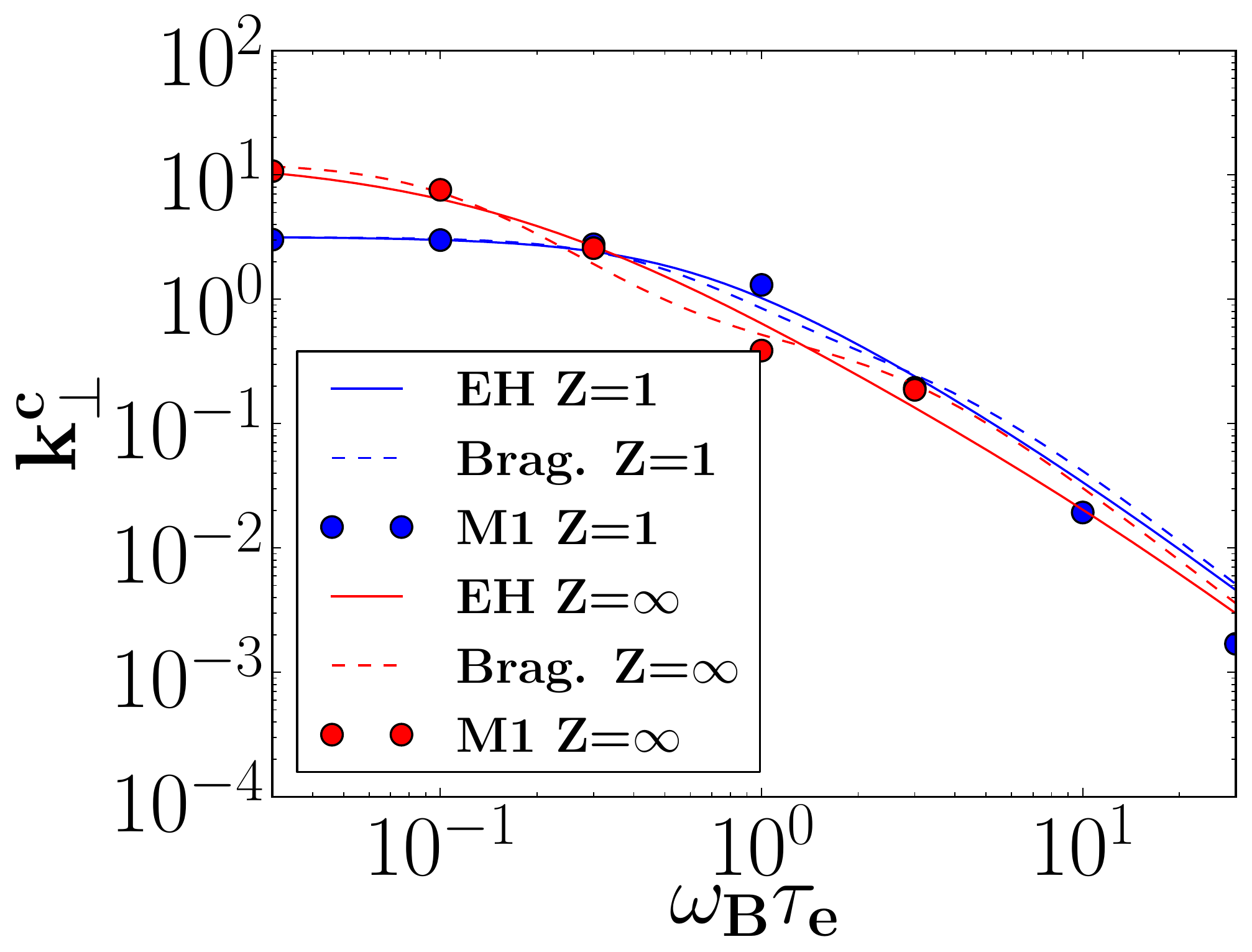}}~
\subfloat[\label{kcross}]{
        \centering
        \includegraphics[width=0.45\columnwidth]{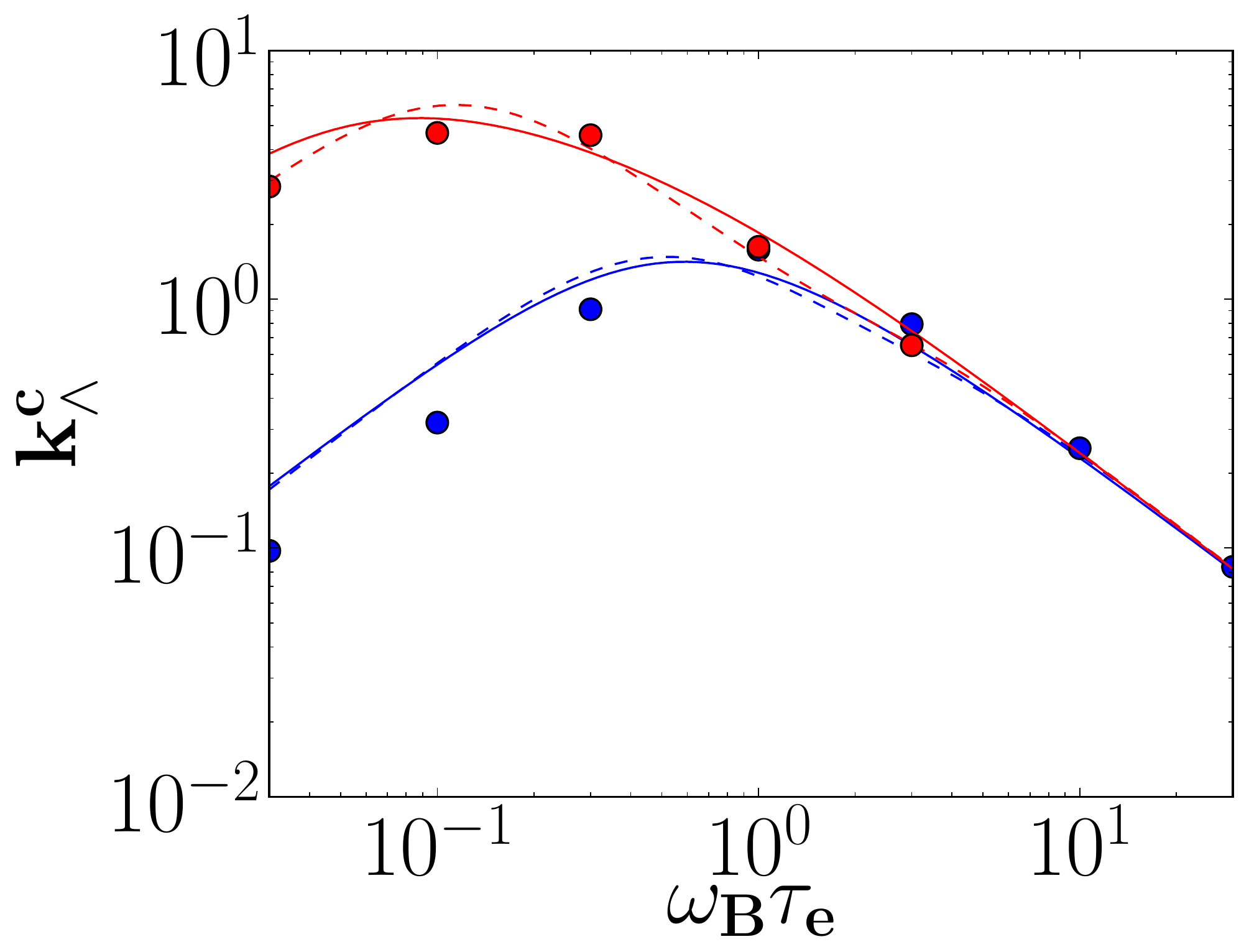}}
    
    \centering
\subfloat[\label{abs_k}]{
        \centering
        \includegraphics[width=0.45\columnwidth]{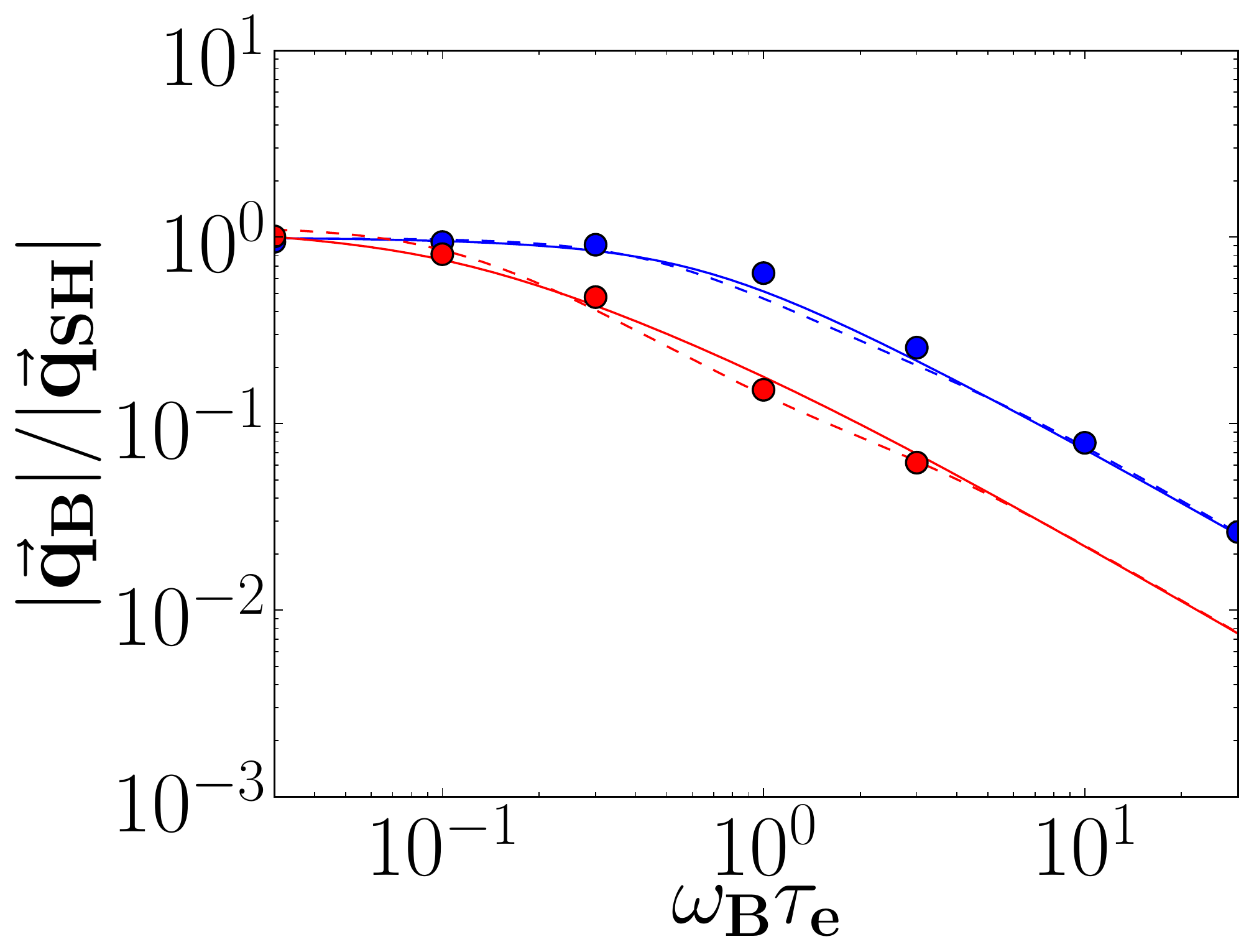}}~
\subfloat[\label{theta_k}]{
        \centering
        \includegraphics[width=0.45\columnwidth]{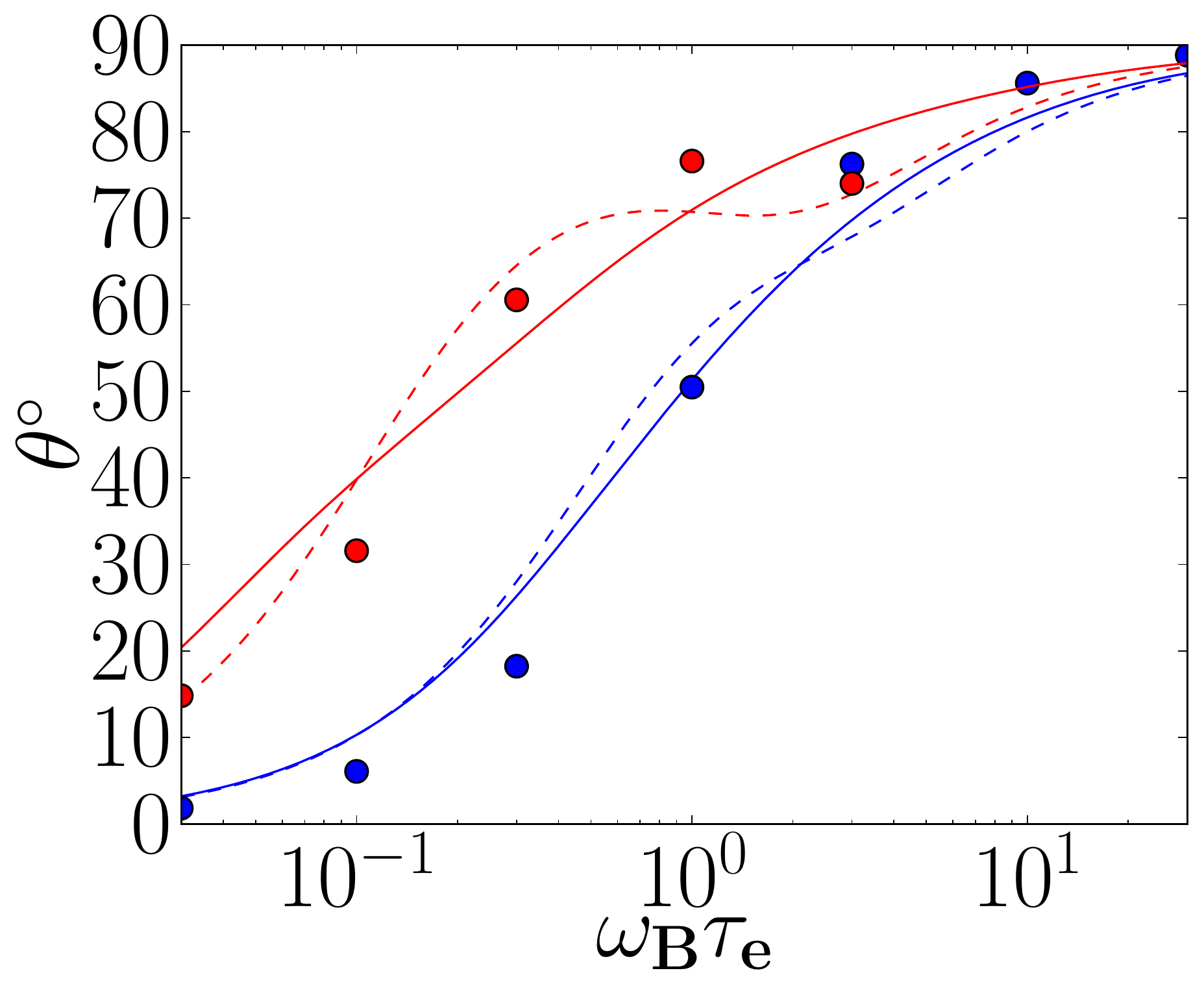}}
    
    \caption{Dependence of dimensionless local thermal conductivities, on the Hall parameter.
    The M1 model (points) is compared to the Braginskii's \citep{Braginskii1965} (continuous line) and the EH theory \citep{epperlein1986plasma} (dashed line), for the conductivity in the direction perpendicular to the magnetic field (a), crossed between the temperature gradient and the magnetic field (b), for the flux reduction (c) and for the flux rotation (d).
    Comparison is presented for $Z=1$ (blue) and $Z\gg1$ (red), $\bar{\bar{k}}^{c}=\bar{\bar{k}}m_{e}/(n_{e}T_{e}\tau_{e})$ is normalized to local quantities.}\label{thermal conductivities}
\end{figure}
The thermal conductivities are compared with the theoretical predictions\citep{Braginskii1965,epperlein1986plasma} in the central plasma region. This comparison is performed in the limit of temperature gradient dominance, as it is demonstrated in Appendix \ref{Demonstration of temperature dominance}.  
The heat flux has the following form
\begin{equation}
\begin{cases}
q_{Bx}\approx-k_{\perp}\frac{\partial}{\partial x}T_{e}
\\
q_{By}\approx-k_{\wedge}\frac{\partial}{\partial x}T_{e}
\end{cases}.\label{caso locale}
\end{equation}

Figure \ref{kperp} presents a dependence of the normalized thermal conductivity $k^{c}_{\perp}= k_{\perp} m_{e}/n_{e}T_{e}\tau_{e}$ versus the Hall parameter, for different values of the ion charge $Z$. 
As $\omega_{B}\tau_{e}$ increases, $k^{c}_{\perp}$ decreases. In accordance to Eq.~\eqref{caso locale}, this corresponds to a decrease of the heat flux in the $x$-direction. In the crossed direction ($y$-axis), shown in Fig.~ \ref{kcross}, the flux (as the conductivity) increases with the magnetization, till a maximum at $\omega_{B}\tau_{e}\approx0.1-1$ (depending on the ion charge) and decreases for larger values of the Hall parameter. 

A description of the flux rotation and limitation, induced by magnetic fields, can be provided by  using a simplified FP model.
Using the Krook collision operator \citep{BGK},  under the assumption of a Lorentz gas and a P1 approximation, the first two angular moments in the system \eqref{first two moments} read
\begin{equation}
\begin{cases}
v\vec{\nabla}\cdot\vec{f}_{1}-\frac{e\vec{E}}{m_{e}v^{2}} \cdot\frac{\partial}{\partial v}\left(v^{2}\vec{f}_{1}\right)=-\nu_{ee}\left(f_{0}-f_{0}^{m}\right)
\\
\frac{v}{3} \vec{\nabla}f_{0}-\frac{e\vec{E}}{3m_{e}}\frac{\partial}{\partial v}f_{0}+\frac{e}{m_{e}c}\vec{f}_{1}\times \vec{B}=-\nu_{ei}\vec{f}_{1}
\end{cases}.
\end{equation}
Inverting the system we recover the first moment \citep{nicolai}:
\begin{equation}
\vec{f}_{1}=\frac{-\lambda_{ei}^{*}}{1+\left(\frac{\omega_{B}\lambda_{ei}^{*}}{v}\right)^{2}}
\left(1+\frac{\omega_{B}\lambda_{ei}^{*}}{v}\hat{b}\times\right)\left(\vec{\nabla}-\frac{e\vec{E}}{m_{e}v}\frac{\partial}{\partial v}\right)
f_{0}.\label{soluzione m1 bgk campi magnetici}
\end{equation}
In the latter equation, the electron-ion MFP $\lambda_{ei}=v_{th}/\nu_{ei}$ has been replaced with the interpolation \citep{nicolai,del2015reduced}
\begin{equation}
\lambda_{ei}^{*}=\frac{Z+0.24}{Z+4.2}\lambda_{ei},\label{fitted nuei}
\end{equation}
in order to account for low ionization number plasmas.
 We interpret $\omega_{B}\lambda_{ei}^{*}/v$ as a kinetic version of the Hall parameter and assume $\omega_{B}\lambda_{ei}^{*}/v\sim \omega_{B}\tau_{e}$, as a leading term in the local limit. According to Eq.~\eqref{heat flux definition}, Eq.~\eqref{soluzione m1 bgk campi magnetici} asserts that the heat flux is reduced by a factor $1+(\omega_{B}\tau_{e})^{2}$, in the direction of the temperature gradient, while, by a factor $\omega_{B}\tau_{e}/[1+(\omega_{B}\tau_{e})^{2}]$, in the  crossed direction, between the temperature gradient and the magnetic field. Thus, the flux is always reduced by the magnetization, in the direction of the temperature gradient, and it is increased for $\omega_{B}\tau_{e}\leq 1$ and reduced for $\omega_{B}\tau_{e}\geq 1$, in the crossed direction.

Figure \ref{kperp} shows that the model M1 agrees  with the theory \citep{epperlein1986plasma}, in the prediction of a flux limitation along the $x$-direction, due to a magnetic field.
In the high-$Z$ limit, the electron-electron collisions become negligible and the M1 model agrees with the Braginskii's and EH theory.

In the low-$Z$ limit, the M1 model weakly departs from the Braginskii's theory for $\omega_{B}\tau_{e}\gtrsim10$.
However, for such a strong Hall parameter, the magnetic field already strongly reduces the heat flux and this difference is not too important.

Also in the $y$-direction, shown in Fig.~\ref{kcross}, the M1 model agrees with the Braginskii's theory in the high-$Z$ limit.
In the low-$Z$ limit it presents some differences when the effect of magnetic fields is weak.
They are due to the inaccuracies of our collision operator, which are more visible for low-Z values.
However, in this limit, the effect of the crossed heat transport is small.

Figures \ref{thermal conductivities} demonstrate that magnetic fields are responsible for two effects: flux limitation and flux rotation.
 By presenting the heat flux in the form of Eq.~\eqref{caso locale}, we define the heat flux limitation due to magnetic fields as\begin{equation}
\frac{|\vec{q}_{B}|}{|\vec{q}_{SH}|}=\frac{\sqrt{k_{\perp}^{2}+k_{\wedge}^{2}}}{k_{SH}}.
\end{equation}
The rotation angle is defined as 
$\theta = \arctan \left({q_{By}}/{q_{Bx}}\right)= \arctan \left({k_{\wedge}^{c}}/{k_{\perp}^{c}}\right)$.

The magnetic field reduces the absolute value of the heat flux, as it is shown in Fig.~\ref{abs_k}.
Our model agrees in the description of the flux limitation, also in the low-$Z$ limit.
The main differences between the model and the theory \citep{epperlein1986plasma} are in the flux rotation angle, shown in Fig.~\ref{theta_k}.
The figure shows that despite of some differences, the M1 model agrees with the theoretical predictions:
the magnetic field tends to rotate the flux and the rotation angle increases with the Hall parameter.
For Hall parameters higher than 1, the rotation angle approaches $90^{\circ}$.

\section{Magnetized nonlocal heat transport}\label{Magnetized nonlocal heat transport}

The magnetized transport is studied in the nonlocal regime, for plasmas characterized by one temperature gradient and periodically modulated temperatures.

\subsection{Magnetized transport along and across the temperature gradient}

 In  this section we analyze a particular case, in which both magnetization and nonlocal effects influence the heat flux. We assume a magnetic field, in the $z$-direction, given by Eq.~\eqref{campo magnetico test}, with a constant Hall parameter $\omega_{B}\tau_{e}=0.5$. Periodic boundary conditions are assumed in the $y$-direction.

\begin{figure*}
    \centering
\subfloat[\label{omegatau 0.5 nl x}]{
        \centering
        \includegraphics[width=0.45\columnwidth]{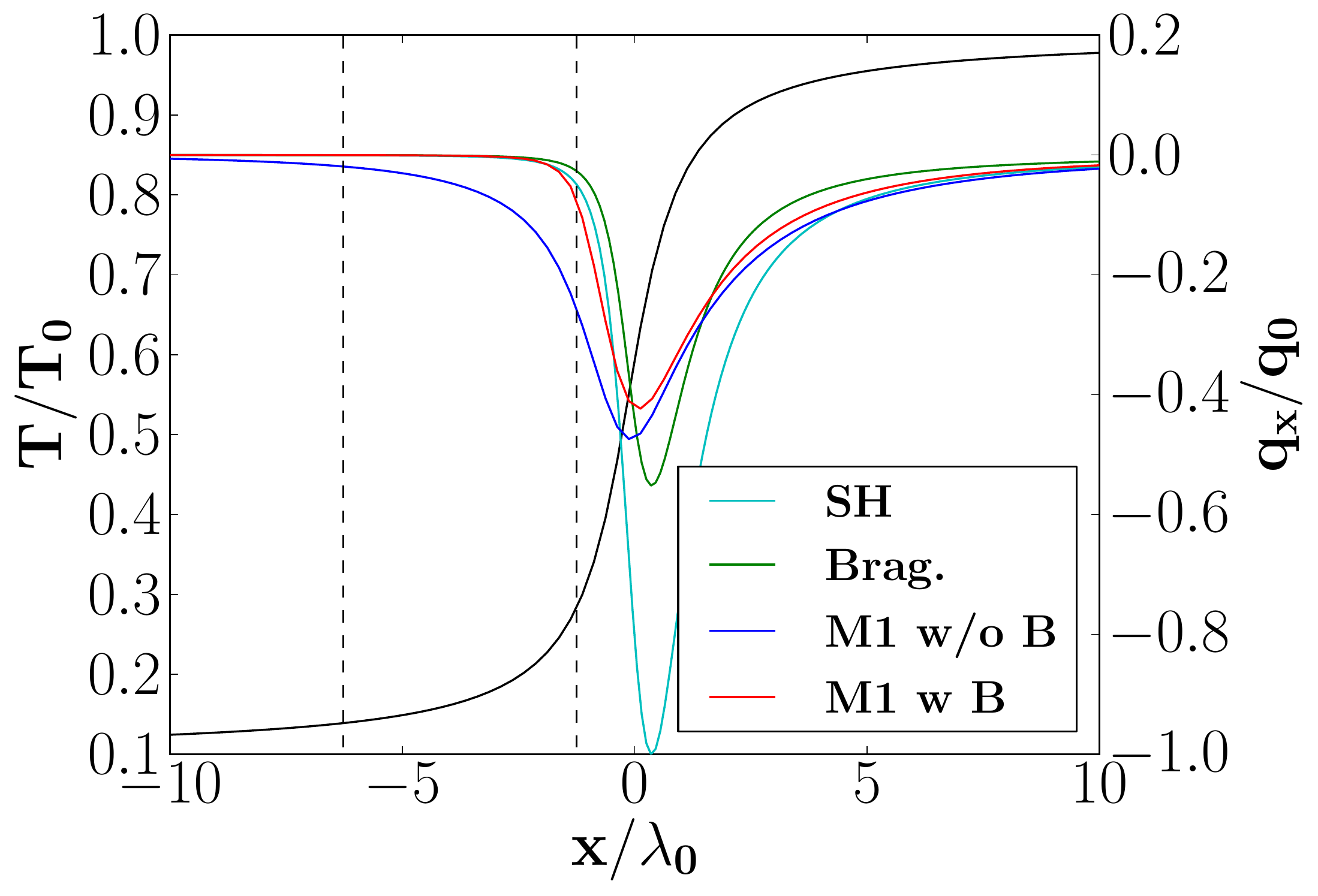}}~ 
\subfloat[\label{omegatau 0.5 nl y}]{
        \centering
        \includegraphics[width=0.45\columnwidth]{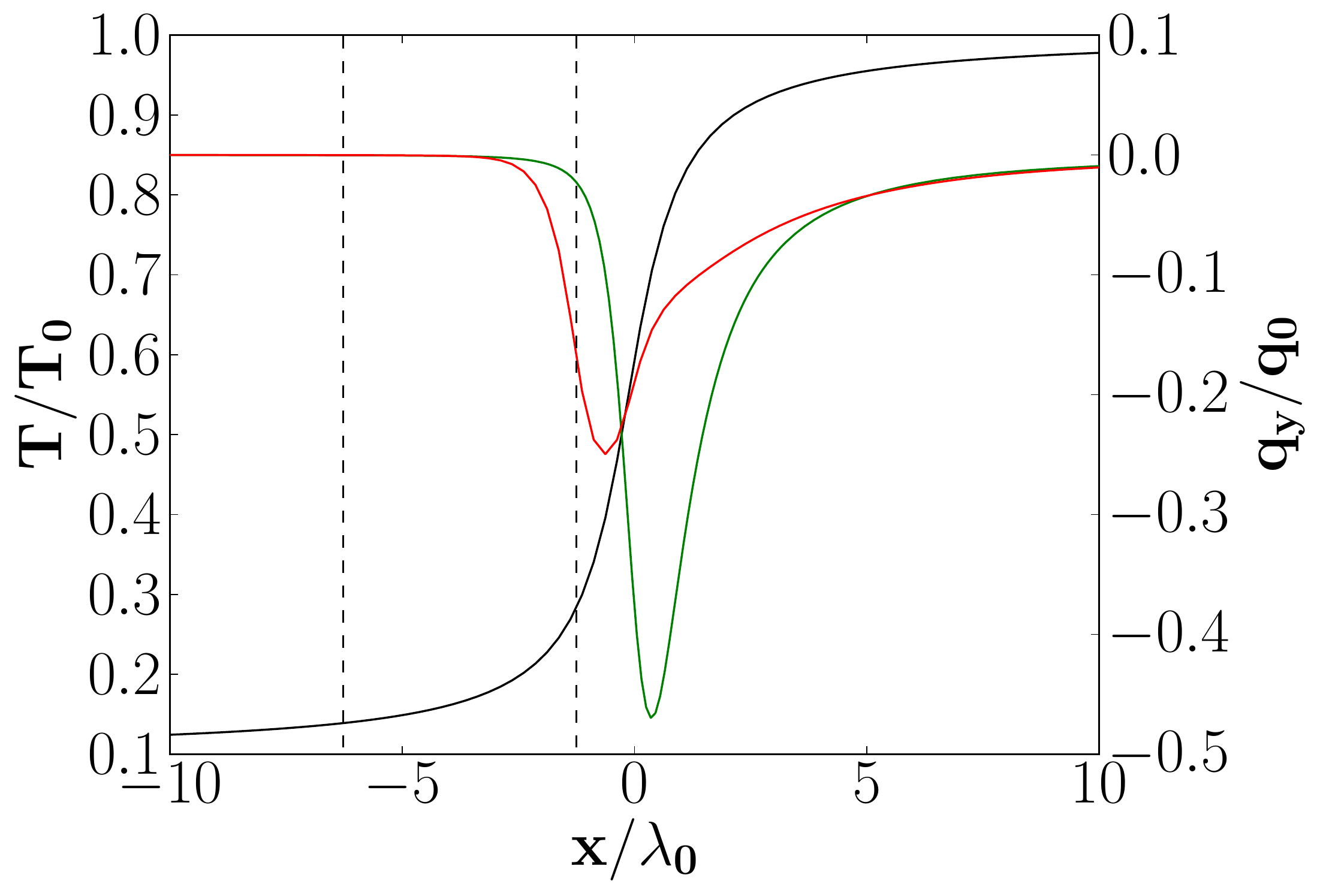}}
    \caption{Nonlocal heat fluxes, versus the $x$-coordinate, for a plasma with a temperature gradient in the $x$-direction, magnetic fields in the $z$-direction and periodic boundary conditions in the $y$-direction. The flux $x$-component is plotted in (a) and the $y$-component in (b). In black is shown the temperature profile, in cyan the local unmagnetized theory ($\omega_{B}\tau_{e}=0$), in green the local magnetized one ($\omega_{B}\tau_{e}=0.5$), in blue the nonlocal M1 model for $\omega_{B}\tau_{e}=0$ and in red the nonlocal M1 model for $\omega_{B}\tau_{e}=0.5$. Vertical lines denote regions where the kinetic analysis is performed.}\label{omegatau 0.5 nl}
    \end{figure*}
We consider a fully-ionized hydrogen plasma of a constant density $n_{e}=10^{23}\, \rm cm^{-3}$.
The temperature profile is given by Eq.~\eqref{temperature atg}  
where $T_{0}=5\;\rm keV$, $T_{1}=0.5\;\rm keV$ and $\delta_{NL}=5\;\mu$m. The fluxes are normalized to the modulus of the maximum SH flux $q_{0}=86\;\rm PW/cm^{2}$  and the length to the maximum MFP $\lambda_{0}=3\sqrt{\frac{\pi}{2}}\frac{T_{0}^{2}}{4\pi n_{e}Ze^{4}\Lambda_{ei}}\approx5.98 \, \mu\rm m$. The degree of nonlocality of the system is $\lambda_{0}/L_{T}=0.67$ and the thermoelectric conduction is negligible.

In order to demonstrate the magnetic field effect on the nonlocal transport, we compare the local and nonlocal models with several magnetization parameters.

Figure \ref{omegatau 0.5 nl x} shows the heat flux along the temperature gradient. The difference between the local and nonlocal models is larger for small $\omega_{B}\tau_{E}$: the SH result is farther from the unmagnetized M1 result than the Braginskii's flux from the magnetized M1, which almost coincide. The magnetic fields in the direction of temperature gradient reduce the nonlocal effect.
On the contrary, the nonlocal effects are stronger in the perpendicular direction, as shown in Fig.~\ref{omegatau 0.5 nl y}. In this case they limit the heat flux and displace it toward the colder region. So, the nonlocal transport reduces the flux rotation due to the magnetic field.

We have seen that the effect of the magnetized nonlocal heat transport is to reduce both nonlocal and magnetic effects. This can be explained by considering the magnetization effectively experienced by an electron with a velocity $v$:
\begin{equation}
\frac{\omega_{B}\lambda_{ei}^{*}}{v}=\sqrt{\frac{2}{9\pi}}\omega_{B}\tau_{e}\left(\frac{v}{v_{th}}\right)^{3}\frac{Z+0.24}{Z+4.2},\label{magn eff}
\end{equation} 
while its effective MFP is $\lambda_{ei}^{*}/[1+(\omega_{B}\lambda_{ei}^{*}/v)^{2}]$. The latter is reduced by magnetic fields for all velocities. A reduction of MFP implies a reduction of the nonlocal parameter $\lambda_{e}/L_{T}$ to $\sim\lambda_{e}/\{L_{T}[1+(\omega_{B}\tau_{e})^{2}]\}$, which necessarily leads to the reduction of nonlocal effects.

\begin{figure}
    \centering
\subfloat[\label{f1p1m1_75.pdf}]{
        \centering
        \includegraphics[width=0.45\columnwidth]{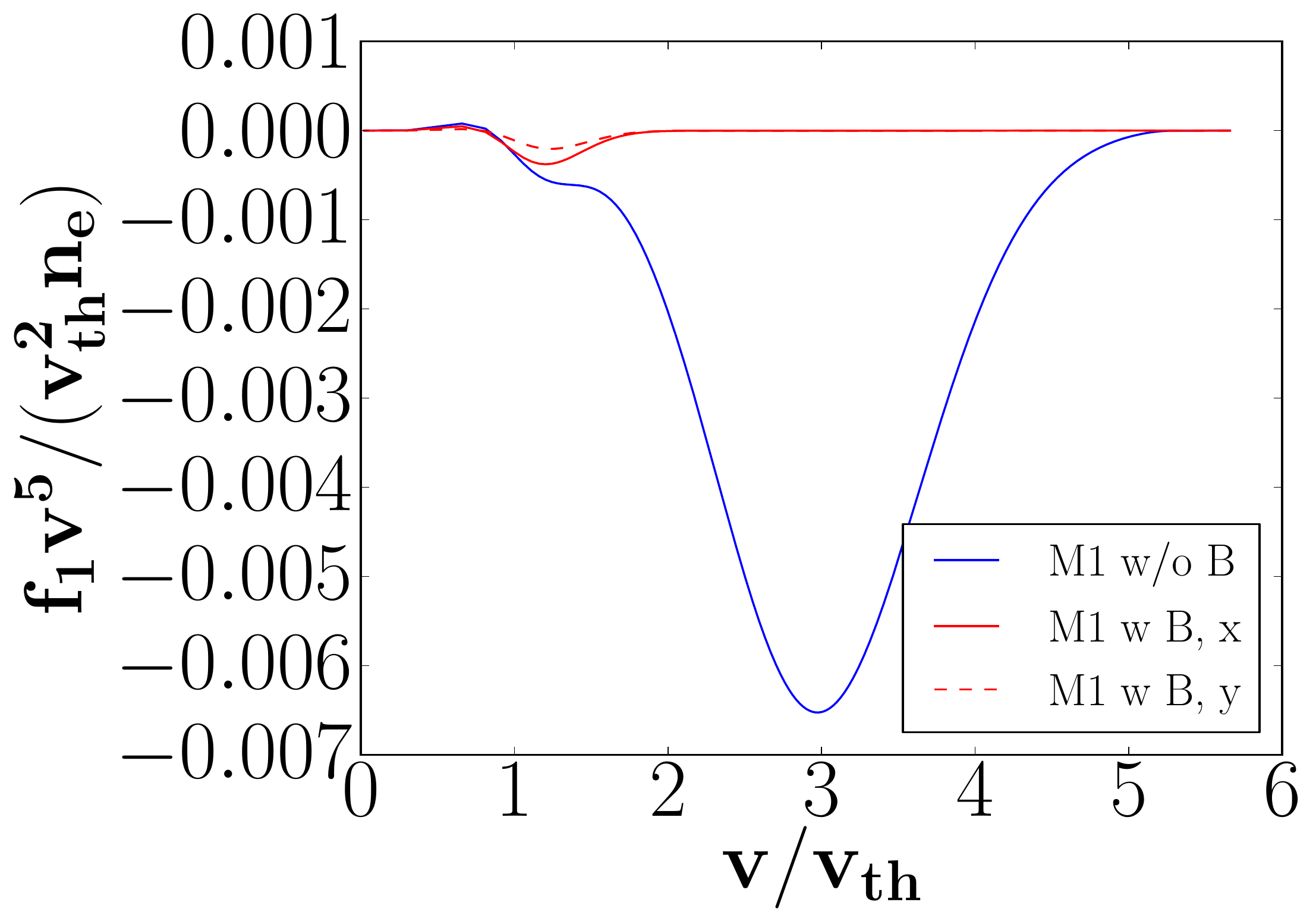}}~
\subfloat[\label{f1p1m1_100.pdf}]{
        \centering
        \includegraphics[width=0.45\columnwidth]{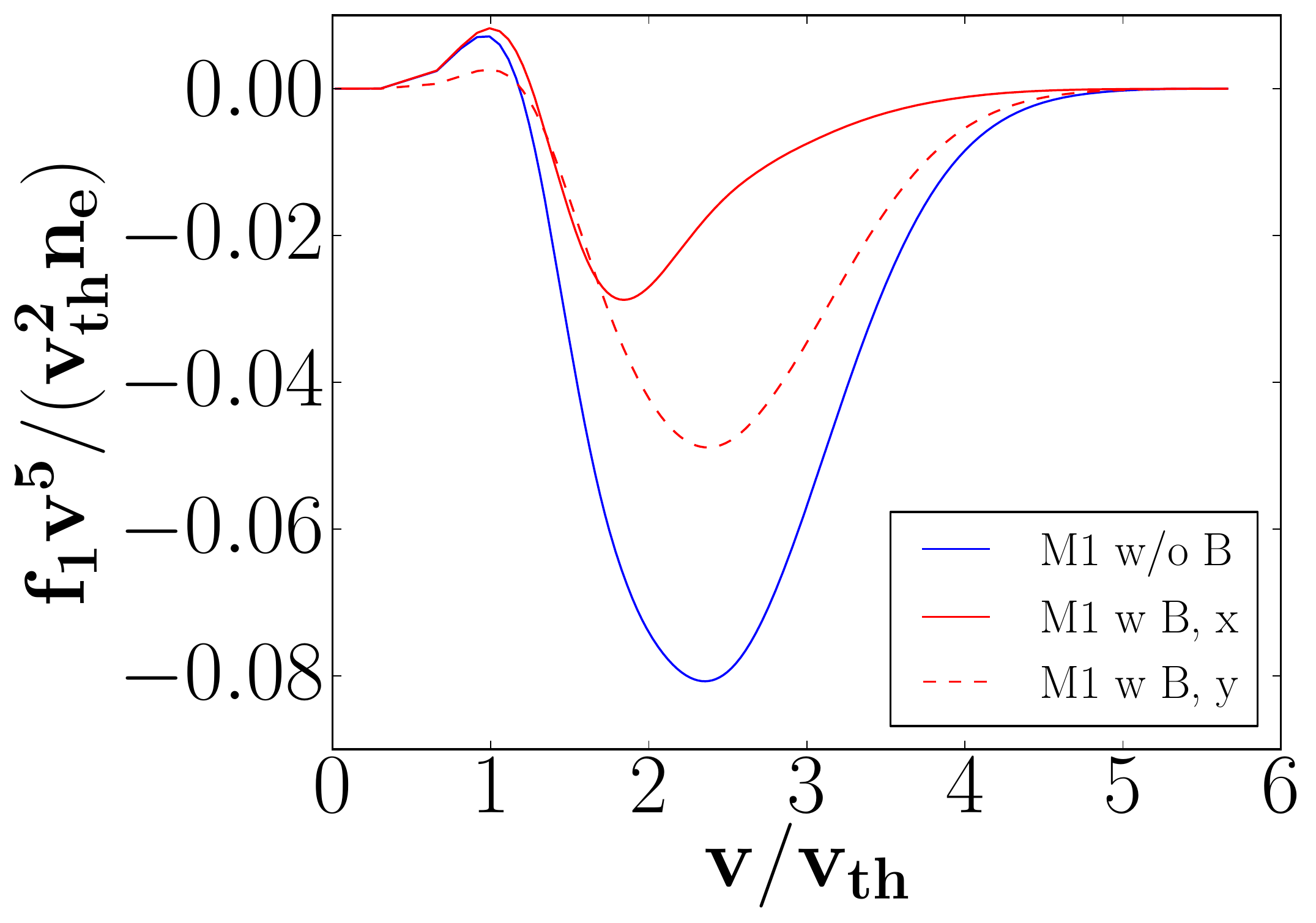}}
    \caption{Magnetized and unmagnetized integrand functions of the heat flux $\propto v^{5}\vec{f}_{1}$, in the cold (a) and central (b) regions, denoted by dashed lines in Fig.~\ref{m1wB_100.pdf}. They are plotted as functions of the velocity modulus. The magnetized flux is split in $f_{1x}$, in the $x$-direction (continuous line) and $f_{1y}$, in the $y$-direction (dotted line).}
    \end{figure}

In Figs.~\ref{f1p1m1_75.pdf} and \ref{f1p1m1_100.pdf} we plot the heat flux integrand functions $\propto v^{5}\vec{f}_{1}$, computed in the cold and central regions, denoted by dashed vertical lines in Fig.~\ref{omegatau 0.5 nl x}.
The local theory predicts the presence of two peaks along the direction of the temperature gradient ${f}_{1x}$. A positive peak at low velocities corresponds to the return current and a negative peak at higher velocities corresponds to the main heat flux. In the central region, variations of ${f}_{1x}$ are small corrections. In the cold region of the unmagnetized plasma a hot nonlocal flux is deposited (higher peak of Fig.~\ref{f1p1m1_75.pdf}), inducing a strong modification  of the heat flux integrand function, compared to the magnetized case, closer to local predictions. This is due to the reduction of the effective MFP by the magnetic field.  The first moment EDF in the crossed direction ${f}_{1y}$, plotted in Figs.~\ref{f1p1m1_75.pdf} and \ref{f1p1m1_100.pdf}, for the magnetized plasma, shows the flux rotation. It becomes dominant for suprathermal velocities, especially close to the central region. The maximum flux contribution is displaced to the lower velocities, in comparison with the value $3.7v_{th}$, predicted by the SH theory.

In the local regime, electrons which transport the heat ($v\approx3.7v_{th}$) experience a magnetization of $\approx3.2\omega_{B}\tau_{e}$. In the nonlocal case, the characteristic velocity of these electrons decreases \citep{epperlein1992nonlocal}. 
In fact, the heat flux is given by a product of the density of electrons $d^{3}v f_{e}$ in the differential volume $d^{3}v$ and their energy flux $\epsilon \vec{v}$. If the gradients become sharper, more electrons with a lower MFP (lower velocities) are present in the front of the gradient. 
This implies a decrease of the effective Hall parameter experienced by these electrons and so a decrease of the heat flux magnetization.  
According to Fig.~\ref{f1p1m1_100.pdf}, in the central region, the heat flux characteristic velocity along the temperature gradient is $||(v_{x}=2v_{th},v_{y}=2.5v_{th})||\approx 3.2v_{th}$, which implies a reduction of the experienced magnetization to $\approx2.1\omega_{B}\tau_{e}$.

\begin{figure}
    \centering
\subfloat[\label{m1woB_75.pdf}]{
        \centering
        \includegraphics[width=0.45\columnwidth]{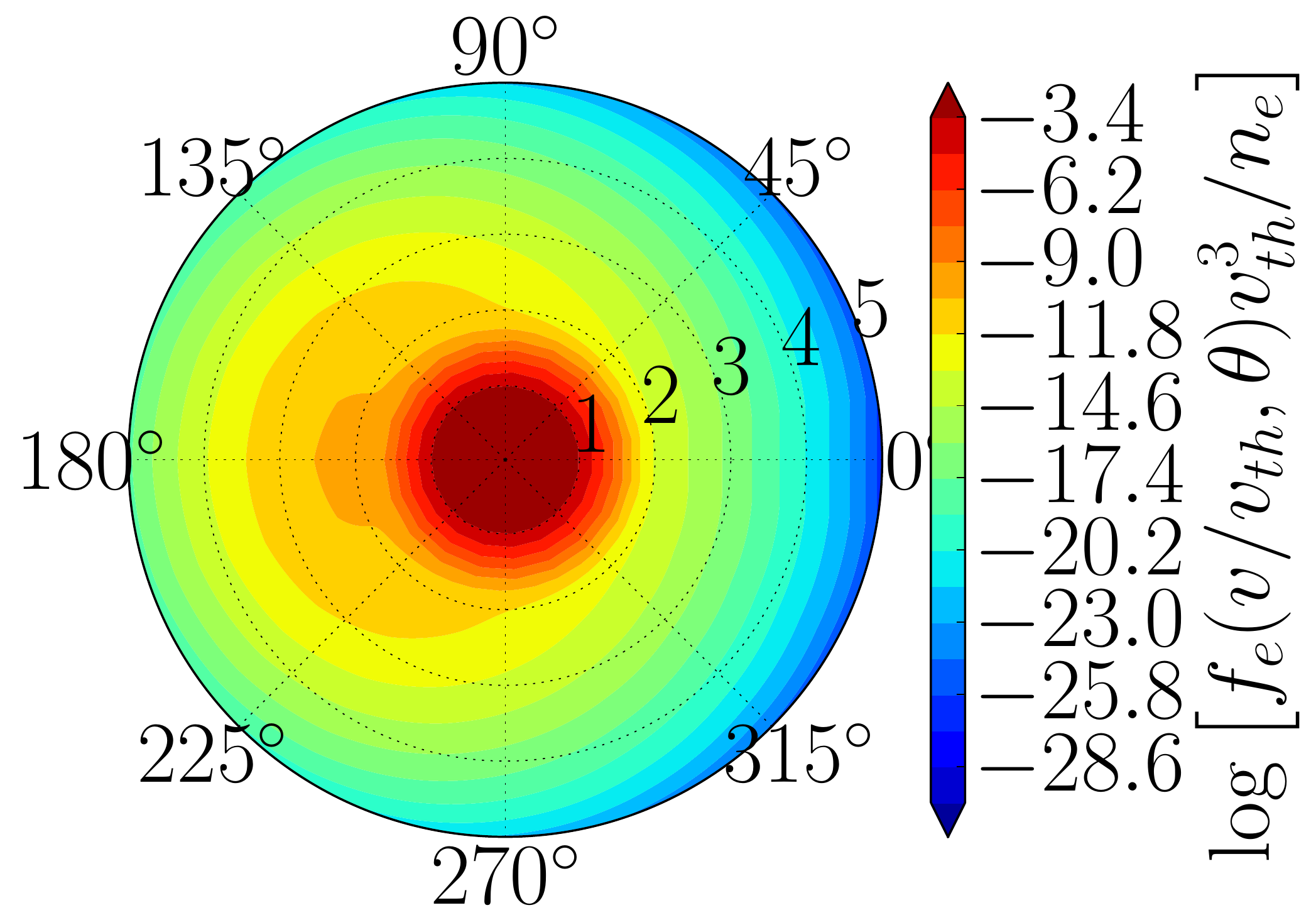}}~ 
\subfloat[\label{m1wB_75.pdf}]{
        \centering
        \includegraphics[width=0.45\columnwidth]{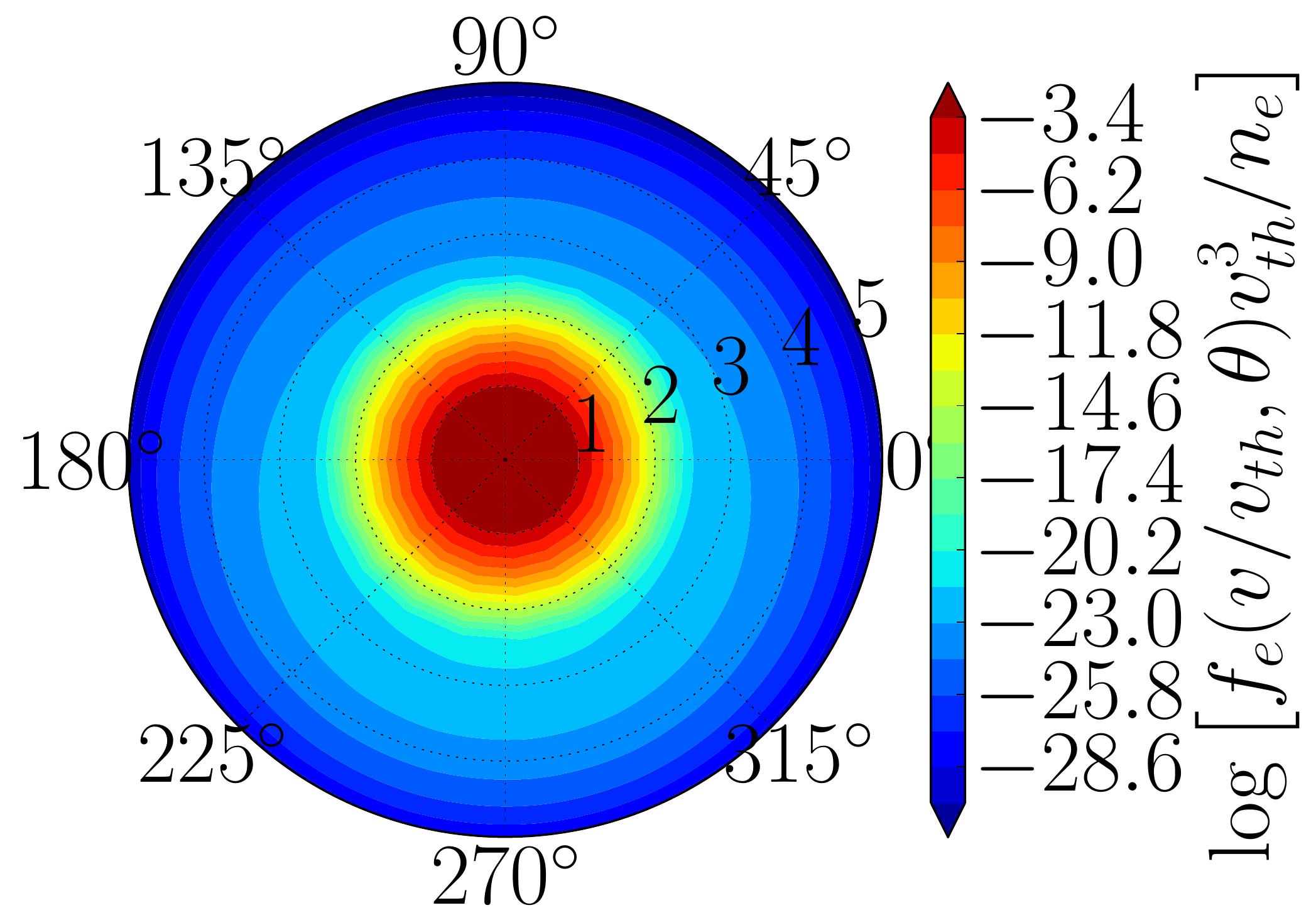}}
        
    \centering
\subfloat[\label{m1woB_100.pdf}]{
        \centering
        \includegraphics[width=0.45\columnwidth]{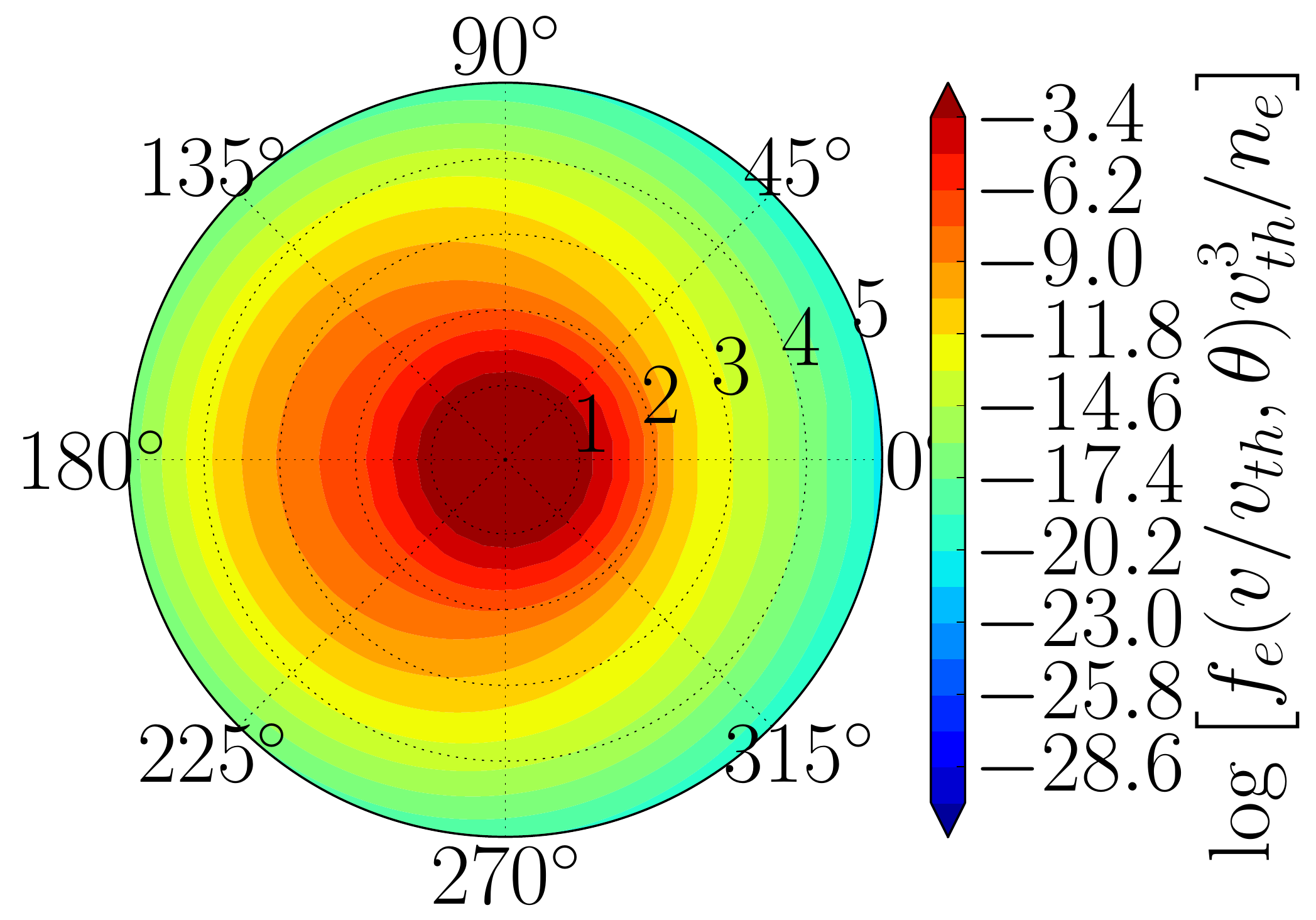}}~
\subfloat[\label{m1wB_100.pdf}]{
        \centering
        \includegraphics[width=0.45\columnwidth]{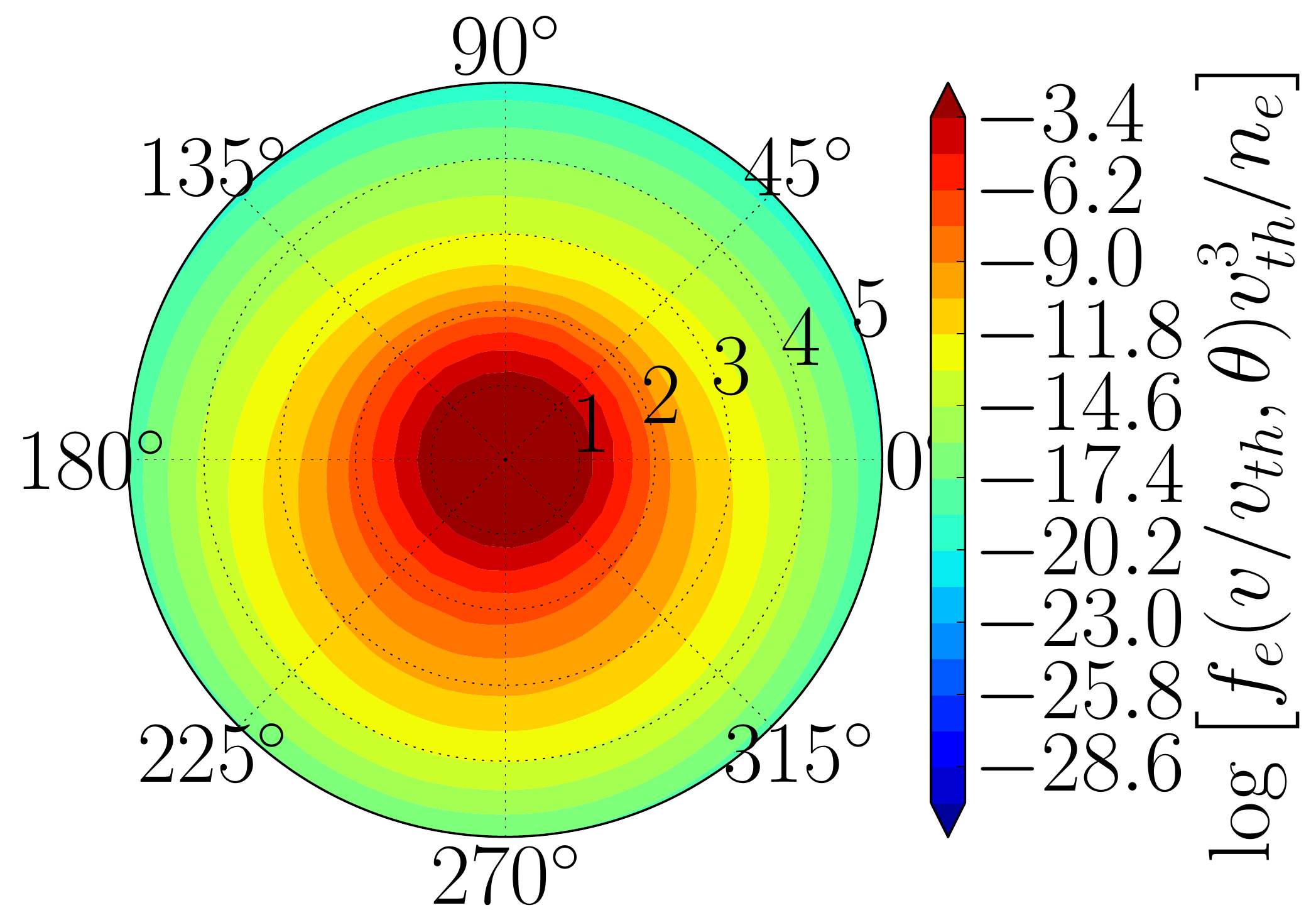}}
     \caption{Unmagnetized ($\omega_{B}\tau_{e}=0$) and magnetized $\omega_{B}\tau_{e}=0.5$ dimensionless logarithm of the EDF, respectively plotted in panels (a) and (b), for the cold region of the plasma and in  (c) and (d), for the central region of the plasma. The analysis is performed in function of the velocity modulus (radial direction of the xy-plane) and of the velocity angle with respect to the temperature gradient (x-axis), in the cold region and in the central region, as indicated by vertical lines, in Fig.~\ref{omegatau 0.5 nl x} and \ref{omegatau 0.5 nl y}.\label{10.pdf}}
    \end{figure}
     The EDFs in the unmagnetized and magnetized regimes, computed with the M1 model are shown in Figs.~\ref{10.pdf} in the region shown in Figs.~\ref{omegatau 0.5 nl x}  and \ref{omegatau 0.5 nl y} by the vertical lines. In the unmagnetized case the EDF presents a direction of anisotropy toward the colder region of the plasma ($180^{\circ}$), due to electrons transporting the heat. Magnetic effects reduce this anisotropy and induce a drift of electrons in the y-direction (Hall effect), affecting the macroscopic transport.  Same effects are visible in the central region, but in this case the Hall current is dominant, due to a stronger electric field. The total EDFs are the ones shown in Figs.~\ref{m1woB_100.pdf} (unmagnetized plasma) and \ref{m1wB_100.pdf}   (magnetized plasma). Note that the Hall parameter experienced by a particle with a velocity $||(v_{x}=2.5v_{th},v_{y}=2.5v_{th})||\approx3.5v_{th}$ is $1.4$, corresponding to the rotation angle  $=\arctan(f_{1y}(2.5v_{th})/f_{1x}(2.5v_{th}))\approx70^{\circ}$ (see Fig.~\ref{theta_k}).

\subsection{Temperature modulation in a magnetized plasma}

Epperlein and Short \citep{epperlein} proposed a test for the study of nonlocal transport.
We extend it to the case of a magnetized plasma.

We consider a fully-ionized beryllium plasma, with a constant density $4.5\times10^{22}\, \rm cm^{-3}$ and the periodic temperature modulation
\begin{equation}
T_{e}(x)=T_{0}+T_{1}\sin(k x),\label{formula ES test}
\end{equation}
with $T_{0}=1\, \rm keV$ and $T_{1}=0.1\, \rm keV$. The magnetic field is applied in the perpendicular direction ($z$), with a constant Hall parameter. Periodic boundary conditions are applied to the $x$ and $y$-directions. The length is normalized by the MFP $\lambda_{0}\approx0.17\;\mu \rm m$.

According to Eq.~\eqref{formula trascuro beta}, thermoelectric effects can be neglected, because the error induced in the local regime, at $x=0$, for $\omega_{B}\tau_{e}=1$, is only $0.04\%$ ($\nu_{e}=2.2\times10^{13}\;\rm s^{-1}$) and even lower in the nonlocal transport regime, which experiences a lower magnetization.

\begin{figure}
    \centering
        \includegraphics[width=0.45\columnwidth]{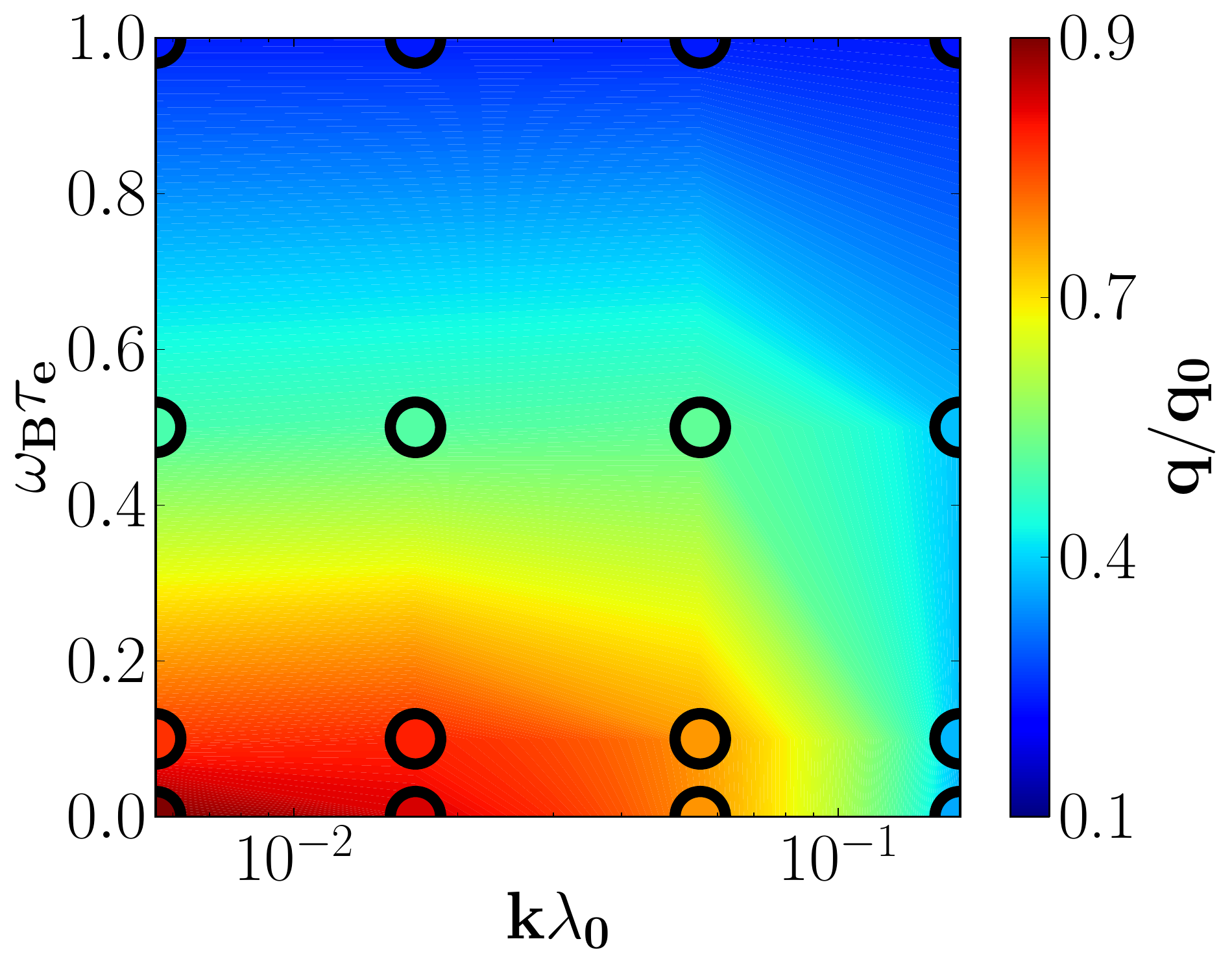}
    \caption{Generalization of the ES test to magnetized plasma. Modulus of the heat flux, normalized to the maximum of the modulus of the SH flux, are given as functions of both nonlocal parameter ($k\lambda_{0}$) and Hall parameter $\omega_{B}\tau_{e}$. Circles indicate the sampled points, while the color background is a linear fit among these points.\label{estest_abs}}
    \end{figure}

The  Hall parameter is varied from the unmagnetized case ($\omega_{B}\tau_{e}=0$), to $\omega_{B}\tau_{e}=1$, for different values of $k\lambda_{0}$, focusing on the flux limitation, where the modulus of fluxes is maximum. The heat flux limitation is shown in Fig.~\ref{estest_abs} by normalizing the heat flux  on the maximum value of the SH flux.

The analysis shows that the nonlocal flux limitation is reduced, increasing the  Hall parameter. This confirms the reduction of nonlocal effects due to the reduction of the effective MFP, as described in the previous section.  
In particular, the flux magnitude is nearly-unmagnetized for $\omega_{B}\tau_{e}=0.1$ and nearly-local for $\omega_{B}\tau_{e}=1$.

Figure \ref{estest_abs} provides a new and simple way to see the combined effect of nonlocal and magnetized transport.

\section{Thermal wave propagation in magnetized plasmas}\label{Relaxation to thermal equilibrium}

The purpose of this section is to describe a propagation of a thermal wave in a magnetized plasma, in the conditions of nonlocal transport. For that, the M1 electron transport model is coupled to the hydrodynamic code CHIC \citep{CHIC}.
We consider a fully-ionized beryllium plasma of constant density $4.5 \times 10^{22}\; \rm cm^{-3}$. The initial electron temperature distribution is given as
\begin{equation}
T_{e}(x,y)=T_{e}(x)e^{-\frac{y^{4}}{y_{max}^{4}}}\left[\theta(x)\left(\frac{x_{\max}}{2x}\right)^{p(y)^{-1}}+\theta(-x)\right],\label{temperatura 2d con formula}
\end{equation}
 with $\theta$ as the Heaviside function, $p(y)=3\exp[y^{2}/(L x_{\max})]$, $L=200 \;\mu \rm m$,  $T_{0}=5 \, \rm keV$, $T_{1}=1 \, \rm keV$, $\delta_{NL}=5\,  \mu \rm{m}$ and $x_{max}=y_{max}=100 \;\mu \rm m$ are the lengths of the target along $x$ and $y$-axes. The initial temperature profile is shown in the background of Figs.~\ref{t0_ot=.pdf}, as a color plot.
 The plasma is simulated with symmetric conditions at the boundaries. 

The magnetization $\omega_{B}\tau_{e}$ is constant in space and time. The magnetic field varies according to
$B_{z}(x,y,t)={m_{e}c}/[{e\tau_{e}(x,y,t)}]\omega_{B}\tau_{e}$.  This choice is not an assumption of the model but just an artifice in order to simplify the analysis and to facilitate a comparison with the Braginskii's theory. In principle, the M1 model is able to deal with any magnetic field configuration.

The electron heat transport generates a thermal wave, which smooths the temperature gradient, till to reach a constant temperature at the thermal equilibrium. 
A temporal evolution of the temperature is given by the heat equation for electrons
\begin{equation}
\frac{3}{2}n_{e}\frac{\partial}{\partial t}T_{e}+\vec{\nabla}\cdot\vec{q}=0,\label{trasporto calore elettronico}
\end{equation}  
assuming the perfect gas approximation. 
 In this example the maximum gradient length is $L_{T}\approx11\;\mu$m. So the characteristic hydrodynamic time is $\tau_{hydro}\sim L_{T}/c_{s}\approx24\;\rm ns$, where the acoustic velocity is $c_{s}=\sqrt{ZT_{0}/m_{i}}=0.46\;\mu$m/ps. On the other hand, the characteristic time of thermal diffusion is $\tau_{diff}\sim n_{e}L_{T}^{2}/\chi\approx1.3$ ps, where $\chi\sim n_{e}v_{th}^{2}\tau_{e}$ is an effective thermal conductivity, with $\tau_{e}=3\sqrt{\frac{\pi}{2}}\frac{m_{e}^{1/2}T_{0}^{3/2}}{4\pi n_{e}Ze^{4}\Lambda_{ei}}\approx0.11 \rm\;ps$ and $v_{th}=\sqrt{T_{e}/m_{e}}\approx30\;\mu$m/ps. Since $\tau_{hydro}>\tau_{diff}$, the hydrodynamic motion can be neglected.

According to Eq.~\eqref{formula trascuro beta}, thermoelectric effects can be neglected, making an error of $\sim 0.3\%$, for $\omega_{B}\tau_{e}=10$.

In what follows we compare the time evolution of the system, using three models for the heat flux $\vec{q}$: the Braginskii's theory, the M1 and the SNB model, in different regimes of magnetization. Equation \eqref{trasporto calore elettronico} is solved for the temperature, injecting at each temporal step the heat flux $\vec{q}$ predicted by the corresponding transport model. The analysis is performed for dimensionless quantities. The temperature is normalized by the initial maximum of the temperature profile $T_{0}=5\;\rm keV$, the space by the maximum MFP $\lambda_{0}\approx3.15 \, \mu\rm m$ and the heat fluxes over the maximum of the magnitude of the initial SH heat flux $q_{0}=40\;\rm PW/cm^{2}$.

\subsection{Analysis of initial conditions}

\begin{figure*}
    \centering
\subfloat[\label{t0_ot=0.pdf}]{
        \centering
        \includegraphics[width=0.3\columnwidth]{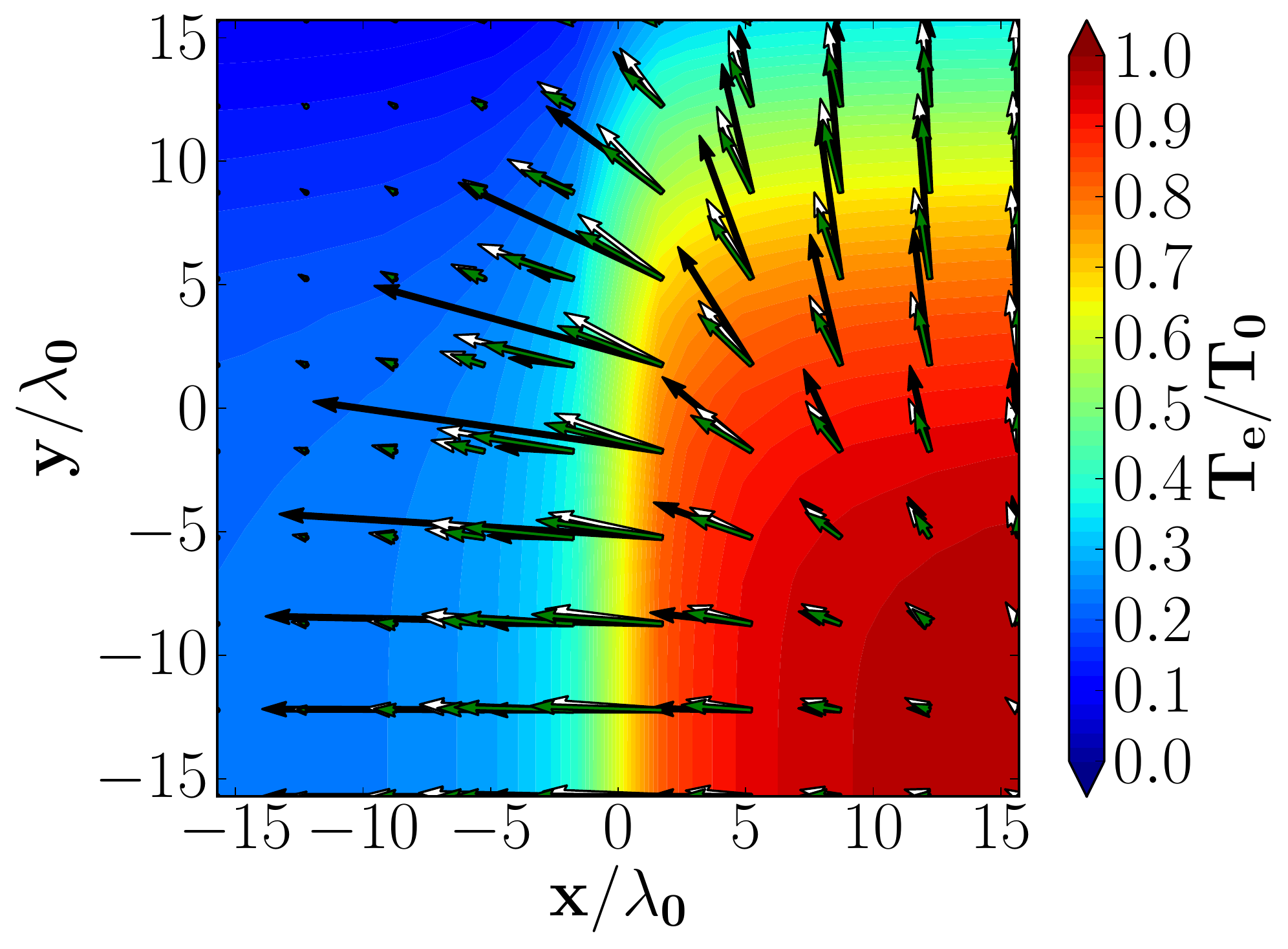}}~%
\subfloat[\label{t0_ot=01.pdf}]{
        \centering
        \includegraphics[width=0.3\columnwidth]{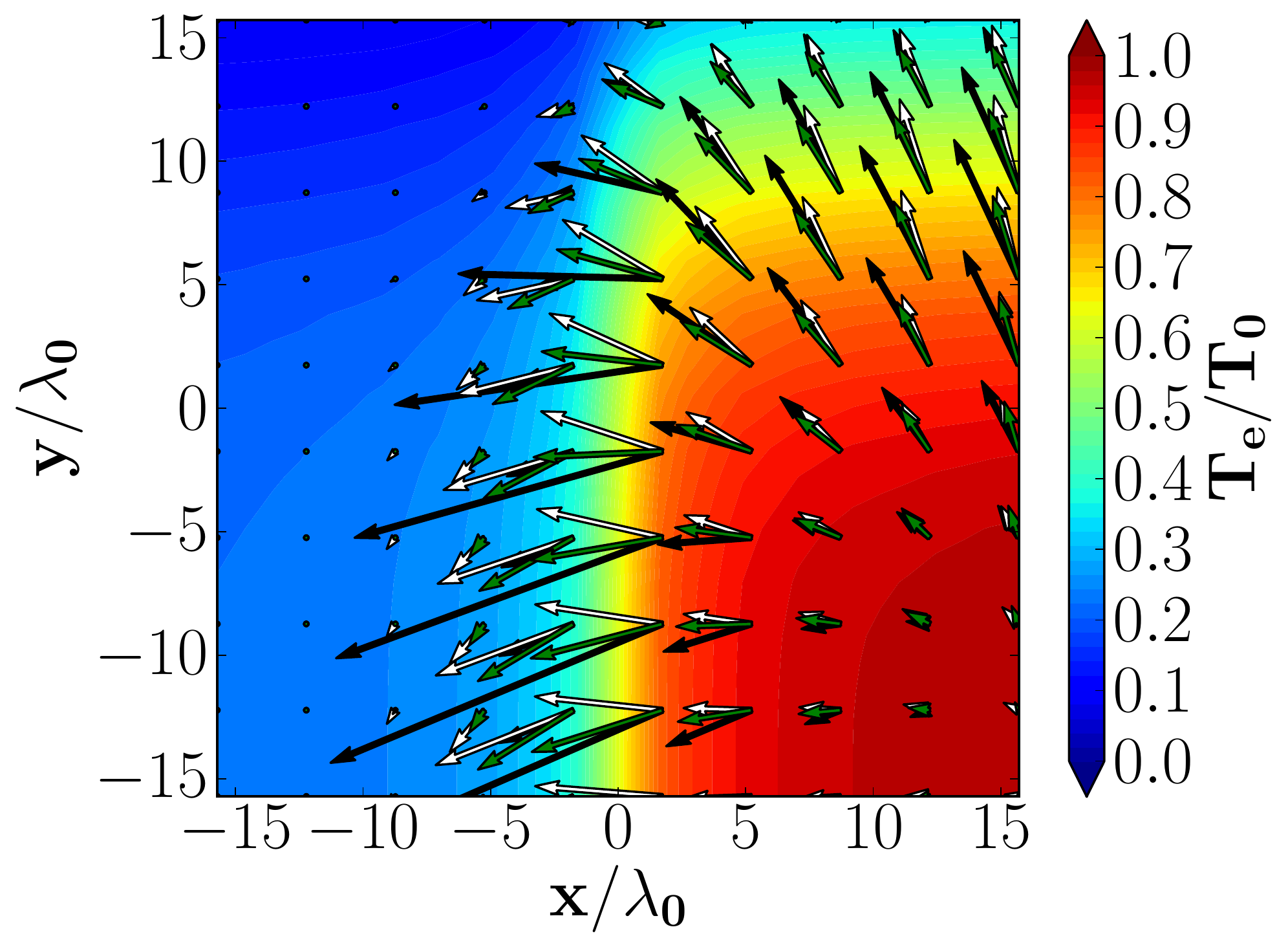}}~%
\subfloat[\label{t0_ot=05.pdf}]{
        \centering
        \includegraphics[width=0.3\columnwidth]{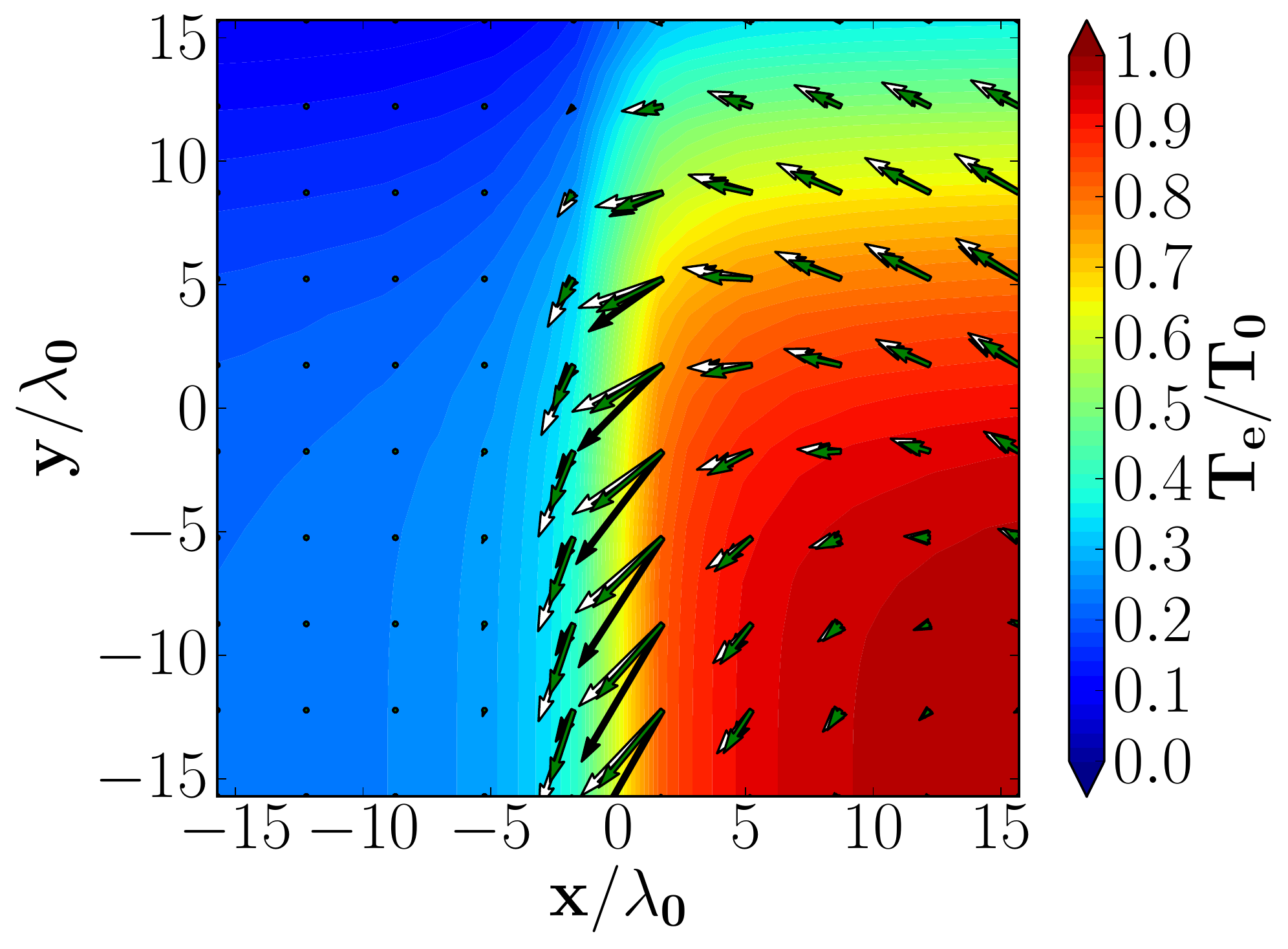}}
        
    \caption{Initial distribution of the electron temperature and heat fluxes, for $\omega_{B}\tau_{e}=0$ (a), $0.1$ (b) and $0.5$ (c). The initial electron temperature profile is shown by the background colors and arrows indicate heat fluxes predicted by different models: the Braginskii's theory (in black), the M1 (in white) and the SNB model (in green).}\label{t0_ot=.pdf}
\end{figure*}

Figures \ref{t0_ot=.pdf} show electron heat fluxes induced by the initial electron temperature profile, appearing in background, as a color plot. Fluxes are computed for increasing values of the Hall parameter (constant in space) and are represented as arrows. 
 The Braginskii's, the M1 and the SNB models are respectively represented in black, white and green.

Figure \ref{t0_ot=0.pdf}, corresponding to the case $\omega_{B}\tau_{e}=0$, shows that the local heat flux is always perpendicular to the temperature gradient. The case of a weak magnetization ($\omega_{B}\tau_{e}=0.1$), reported in Fig.~\ref{t0_ot=01.pdf}, shows a local flux rotation, respect to the unmagnetized case. Increasing the magnetization ($\omega_{B}\tau_{e}>0.1$), we can see another effect: a reduction of the local flux. These two effects become dominant in  Fig.~\ref{t0_ot=05.pdf}, for $\omega_{B}\tau_{e}=0.5$. Finally, for a high degree of magnetization ($\omega_{B}\tau_{e}\gg1$),  the flux is suppressed.

Figure \ref{t0_ot=0.pdf} shows the main effects of unmagnetized nonlocal transport, described in the previous chapter. A flux limitation can be seen near $x\approx 0$, the preheating near $x\approx -5$ and the rotation is negligible. The M1 and the SNB models  quite well agree  in the description of nonlocal fluxes.

When a small magnetization is imposed ($\omega_{B}\tau_{e}=0.1$), the direction of heat flux deviates from the direction of the temperature gradient, due to magnetic fields. The flux rotation, in nonlocal models, appears weaker compared to the local case. This qualitative effect can be seen in both M1 and SNB models; however, they slightly differ in the value of this rotation. The M1 model corresponds to a smaller one, compared to the SNB model. For both models, the flux reduction, due to magnetic fields, becomes important for $\omega_{B}\tau_{e}=0.5$. This confirms that nonlocal fluxes are less affected by magnetic fields, compared to their local counterparts. This conclusion is in accordance with the studies presented in Fig.~6 of Ref.~\citep{brantov2003linear}.

For a strong magnetization ($\omega_{B}\tau_{e}=0.5$), nonlocal models approach the local predictions. These models superpose in the case of flux suppression \citep{phddario} ($\omega_{B}\tau_{e}\gg1$). This is in accordance with the results plotted in Fig.~\ref{estest_abs}, where $\lambda_{0}/L_{T}=0.1k\lambda_{0}$. This behavior is due to a reduction of the MFP, induced  by magnetic fields \citep{nicolai}.

\subsection{Temporal evolution}

\begin{figure}
    \centering
\subfloat[\label{2d_sh.pdf}]{
        \centering
        \includegraphics[width=0.45\columnwidth]{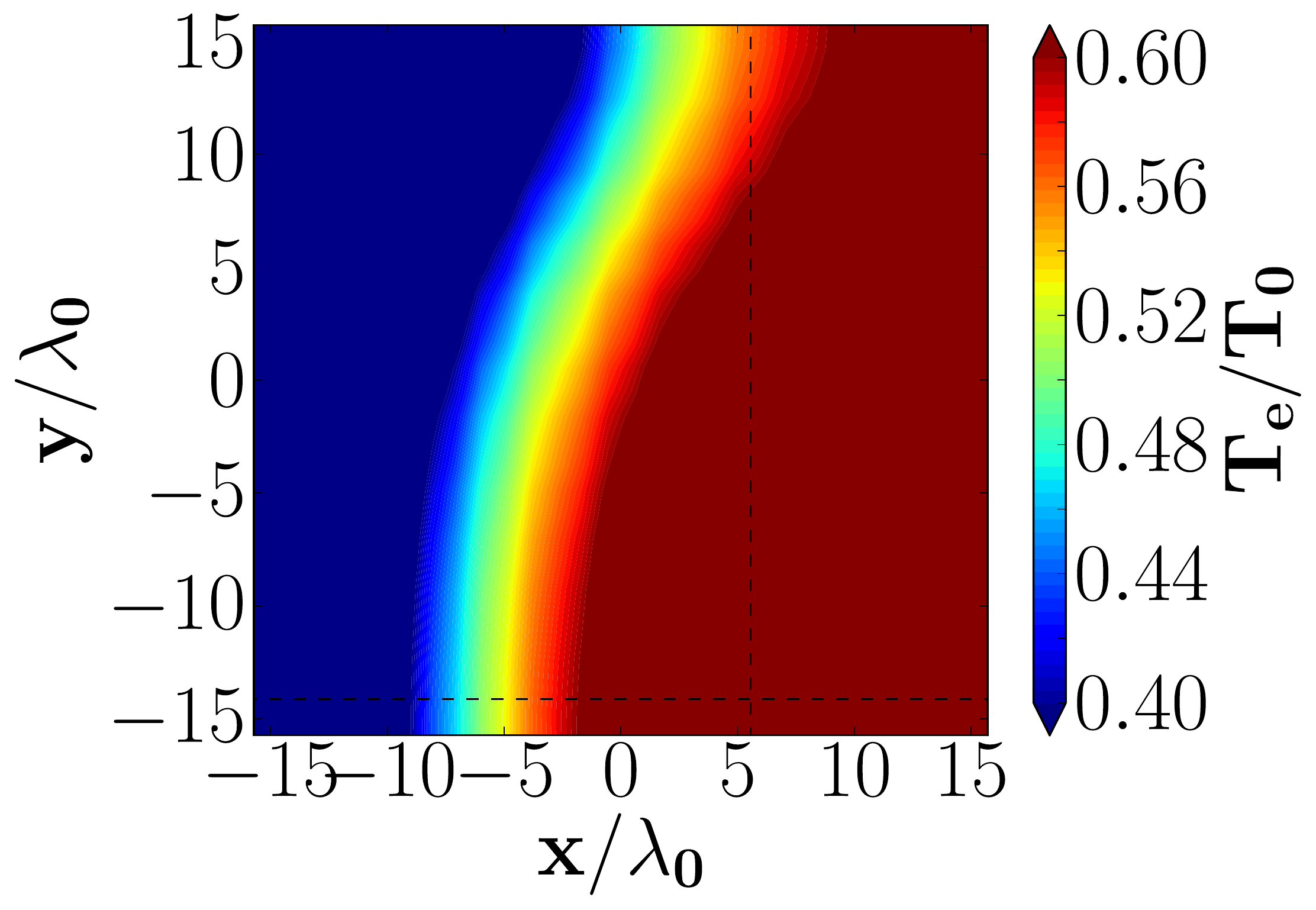}}~ 
\subfloat[\label{2d_m1.pdf}]{
        \centering
        \includegraphics[width=0.45\columnwidth]{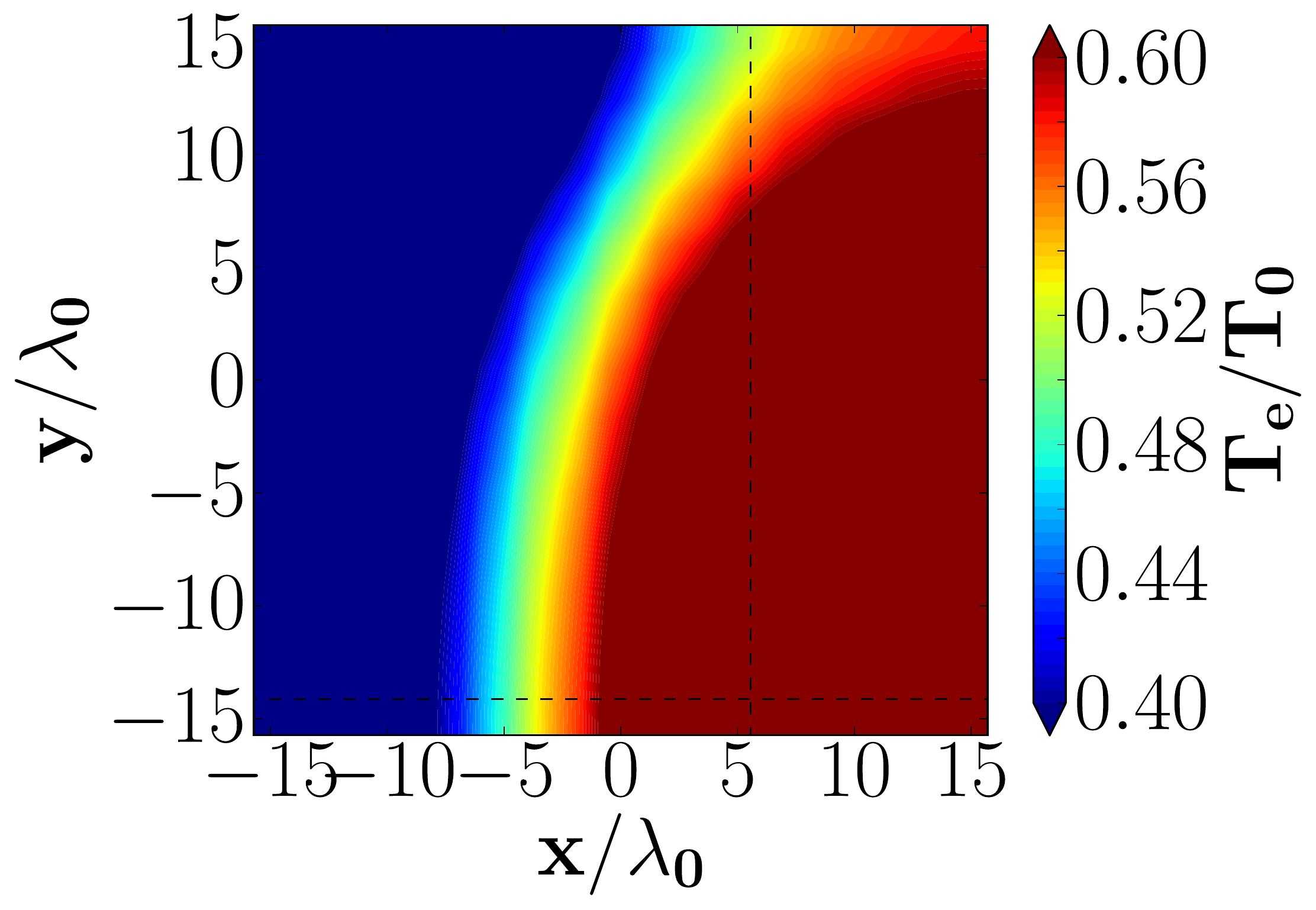}}
    
\subfloat[\label{2d_brag_ot=01.pdf}]{
        \centering
        \includegraphics[width=0.45\columnwidth]{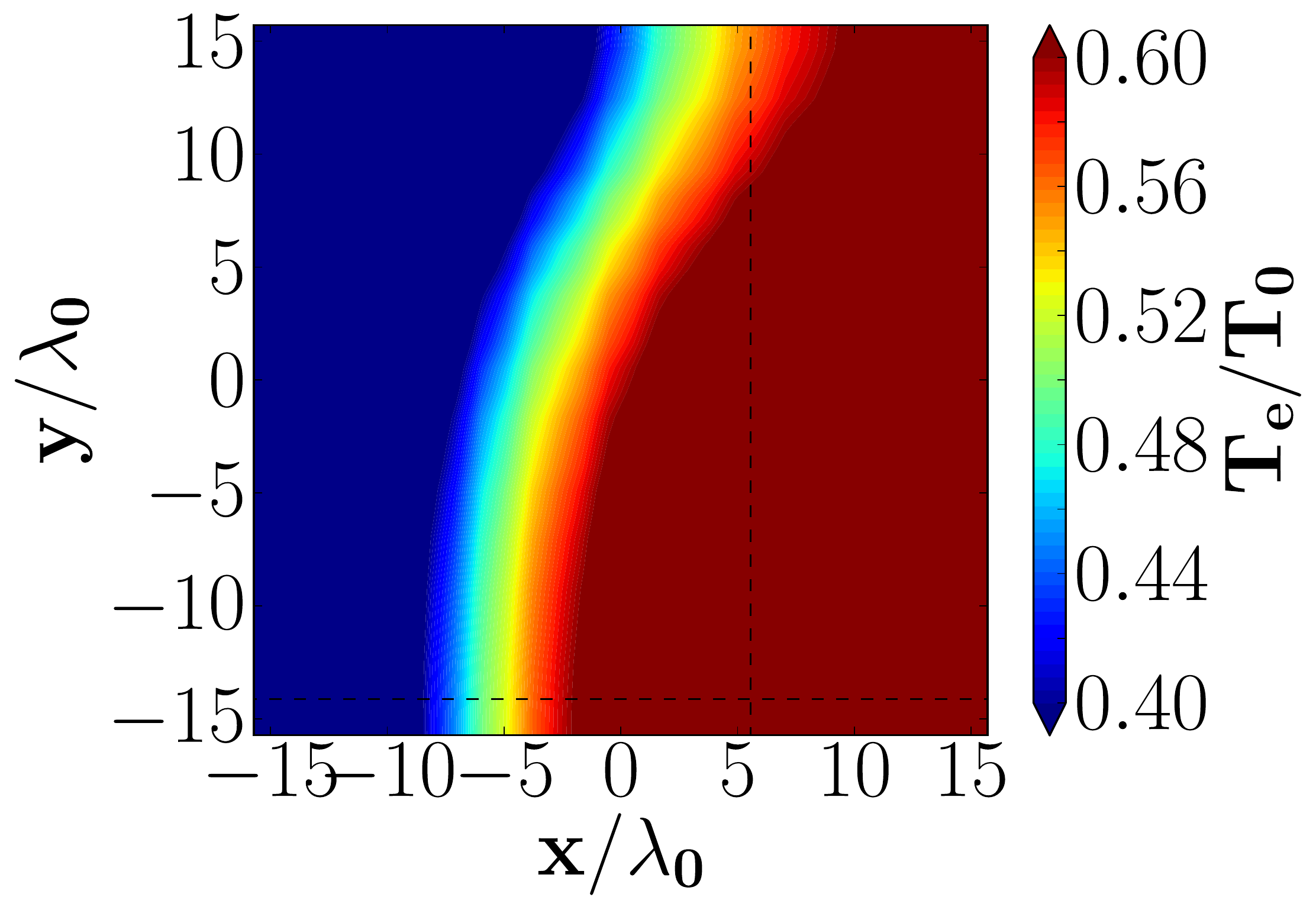}}~
\subfloat[\label{2d_m1_B_ot=01.pdf}]{
        \centering
        \includegraphics[width=0.45\columnwidth]{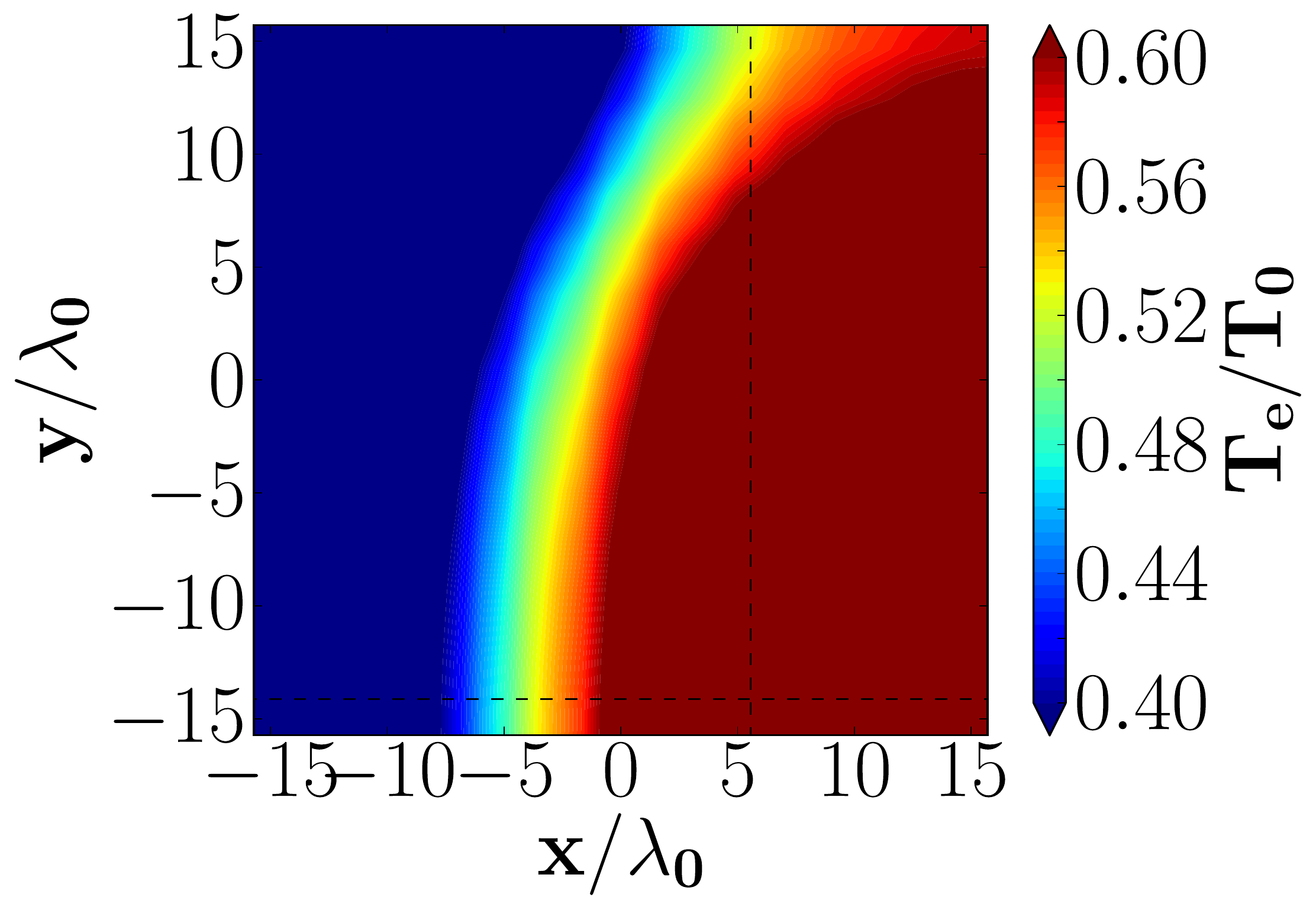}}
    
\subfloat[\label{2d_brag_ot=05.pdf}]{
        \centering
        \includegraphics[width=0.45\columnwidth]{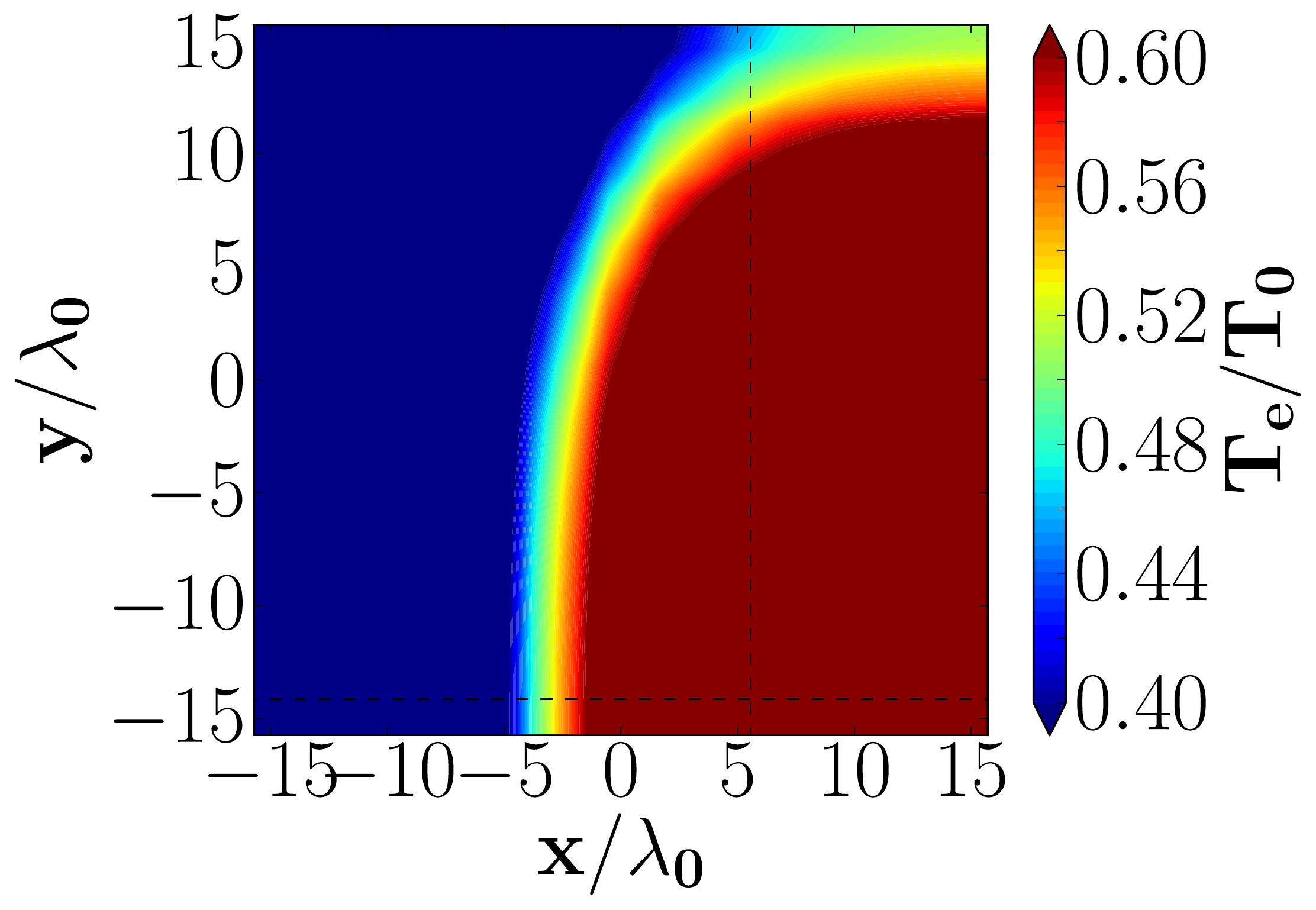}}~ 
\subfloat[\label{2d_m1_B_ot=05.pdf}]{
        \centering
        \includegraphics[width=0.45\columnwidth]{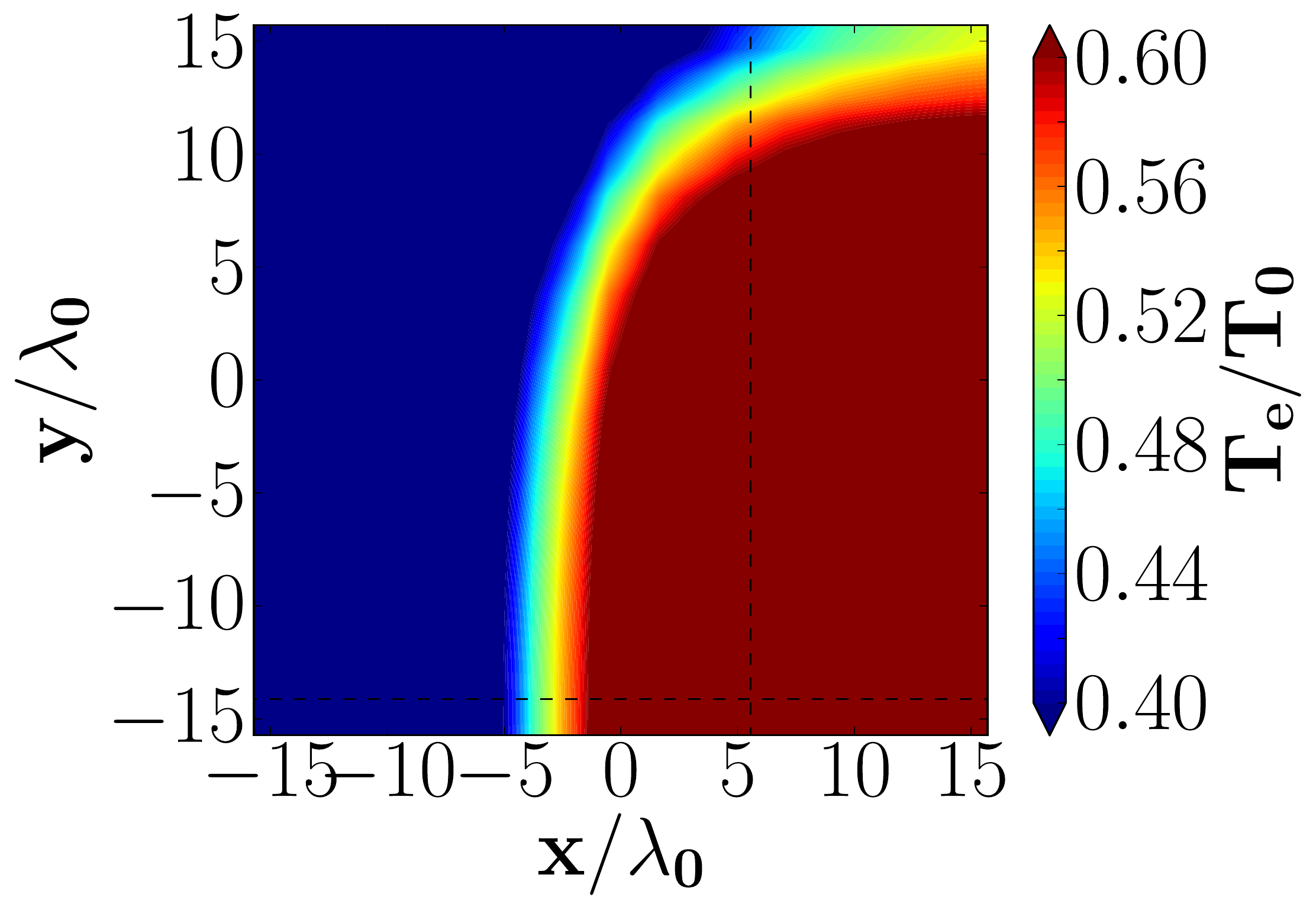}}
    
    \caption{Plasma electron temperature distribution, after $45\tau_{e}$. 
     Dashed lines indicate the regions where line-outs have been performed.
     Panels (a) and (b) correspond to the case without magnetic field, respectively using SH and M1 models. Panels (c) and (d) correspond to $\omega_{B}\tau_{e}=0.1$, respectively using Braginskii's and M1 models. Panels (e) and (f) to  $\omega_{B}\tau_{e}=0.5$,  respectively using the Braginskii's and M1 models.  The SNB model has not been reported since, in this case, the differences from the M1 model are small and not easily visible in two dimensions. 
     \label{2d_m1_B_ot=}}
\end{figure}

Equation \eqref{trasporto calore elettronico} conserves the total energy in the simulation box, regardless of the model used.

Electrons, in the hot region, have a velocity $\sim v_{th}=\sqrt{T_{0}/m_{e}}\approx3\times10^{9}\;\rm cm/s$. The maximum distance of  the simulation box is $45\lambda_{0}$ (diagonal direction). In the unmagnetized regime, electrons take $45\tau_{e}\approx5\;\rm ps$ to travel along the simulation box. At this time, the differences between models are maximum. In subsequent times, they are reduced, since temperature gradients become smoother. 

Figures 
\ref{2d_sh.pdf} and \ref{2d_m1.pdf}
 show respectively the Braginskii's and the M1  temperature predictions, after $45\tau_{e}$, in the unmagnetized regime. The local heating is more efficient, because of the flux limitation which characterizes nonlocal models. The local flux smooths the temperature gradients more than in nonlocal case. The SNB model provides results very close to M1. 

\begin{figure*}
    \centering
\subfloat[\label{cut_y.eps}]{
        \centering
        \includegraphics[width=0.45\columnwidth]{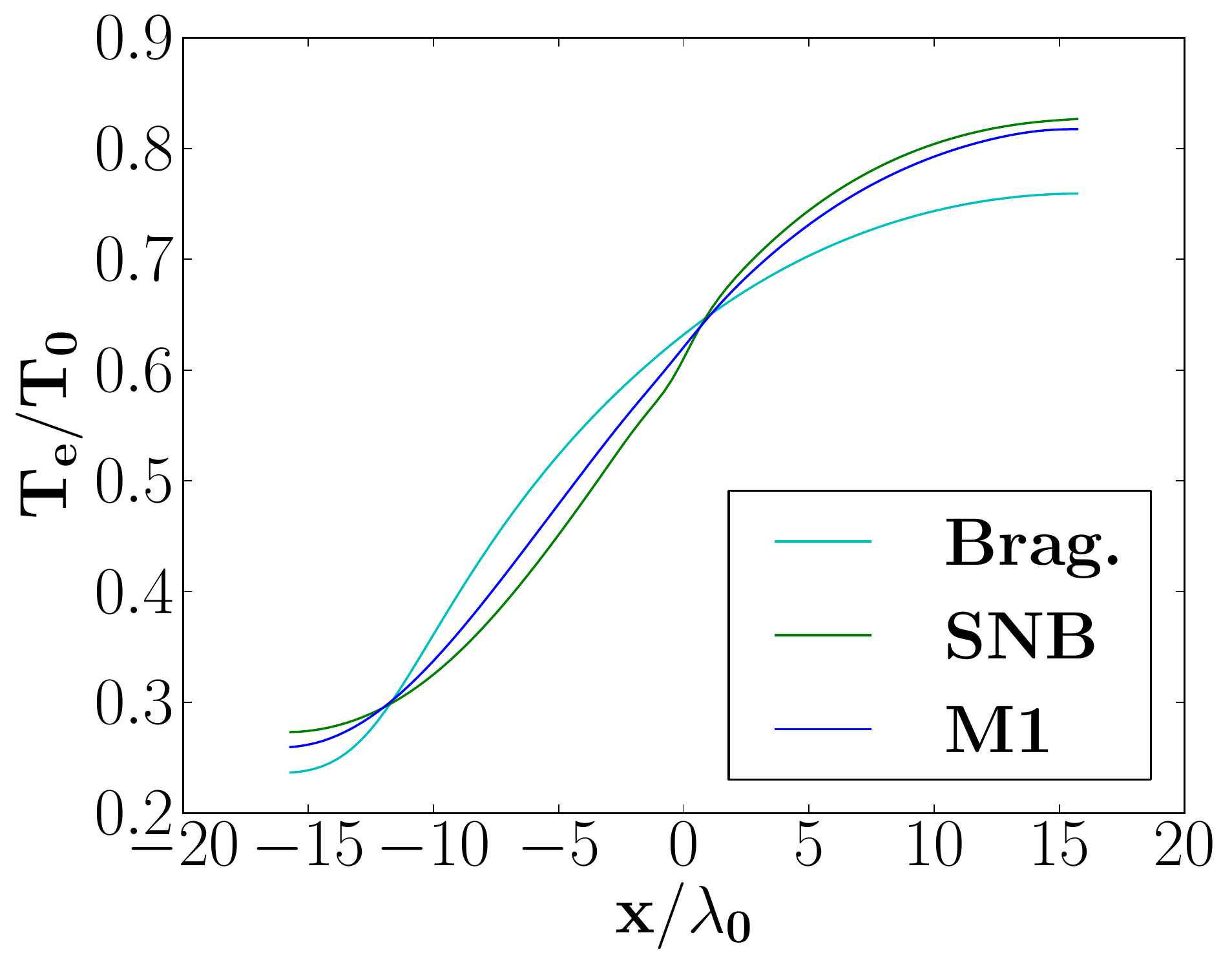}}~ 
    \subfloat[\label{cut_x.eps}]{
            \centering
        \includegraphics[width=0.45\columnwidth]{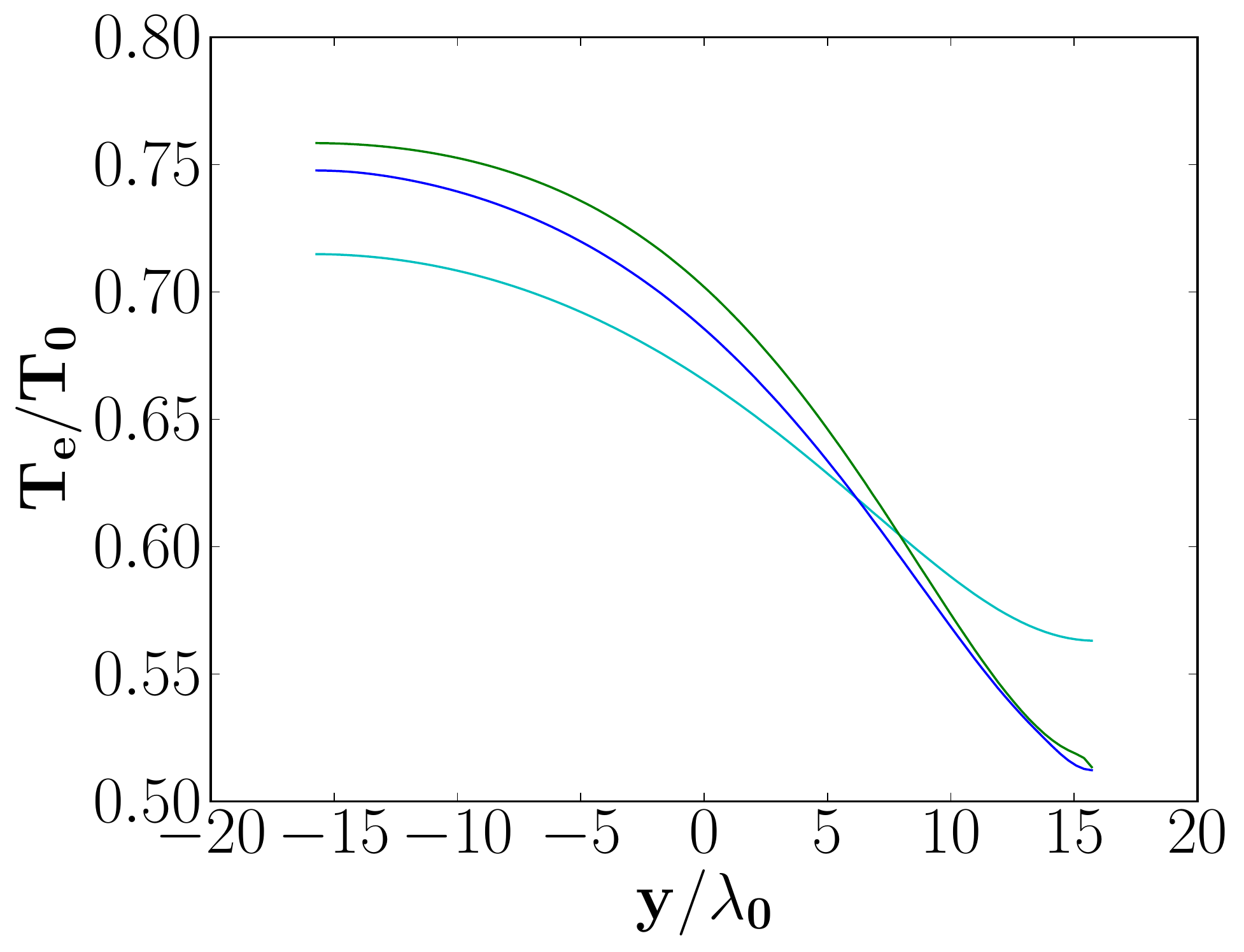}}~

\subfloat[\label{cut_y_ot=01.eps}]{
        \centering
        \includegraphics[width=0.45\columnwidth]{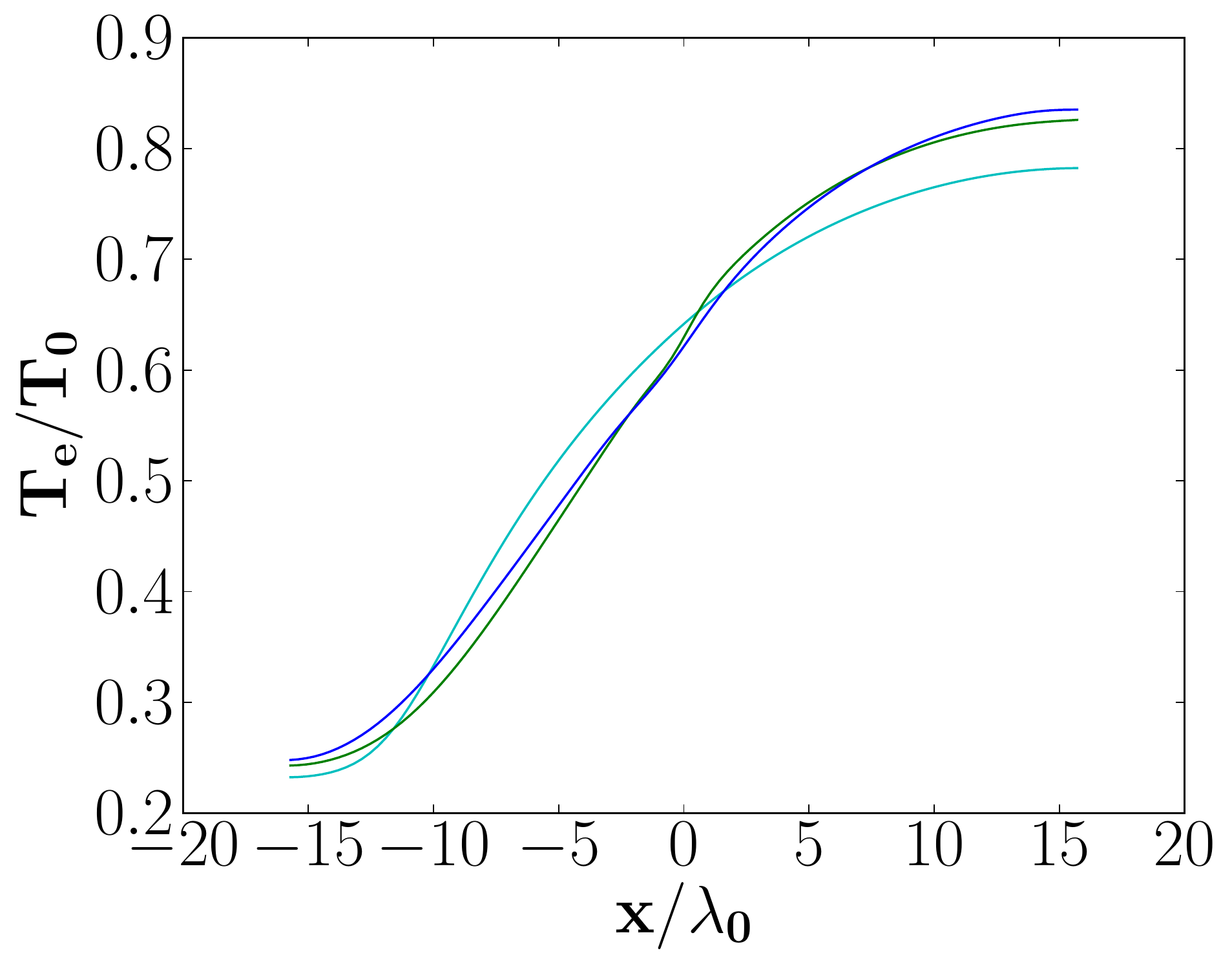}}~ 
\subfloat[\label{cut_x_ot=01.eps}]{
        \centering
        \includegraphics[width=0.45\columnwidth]{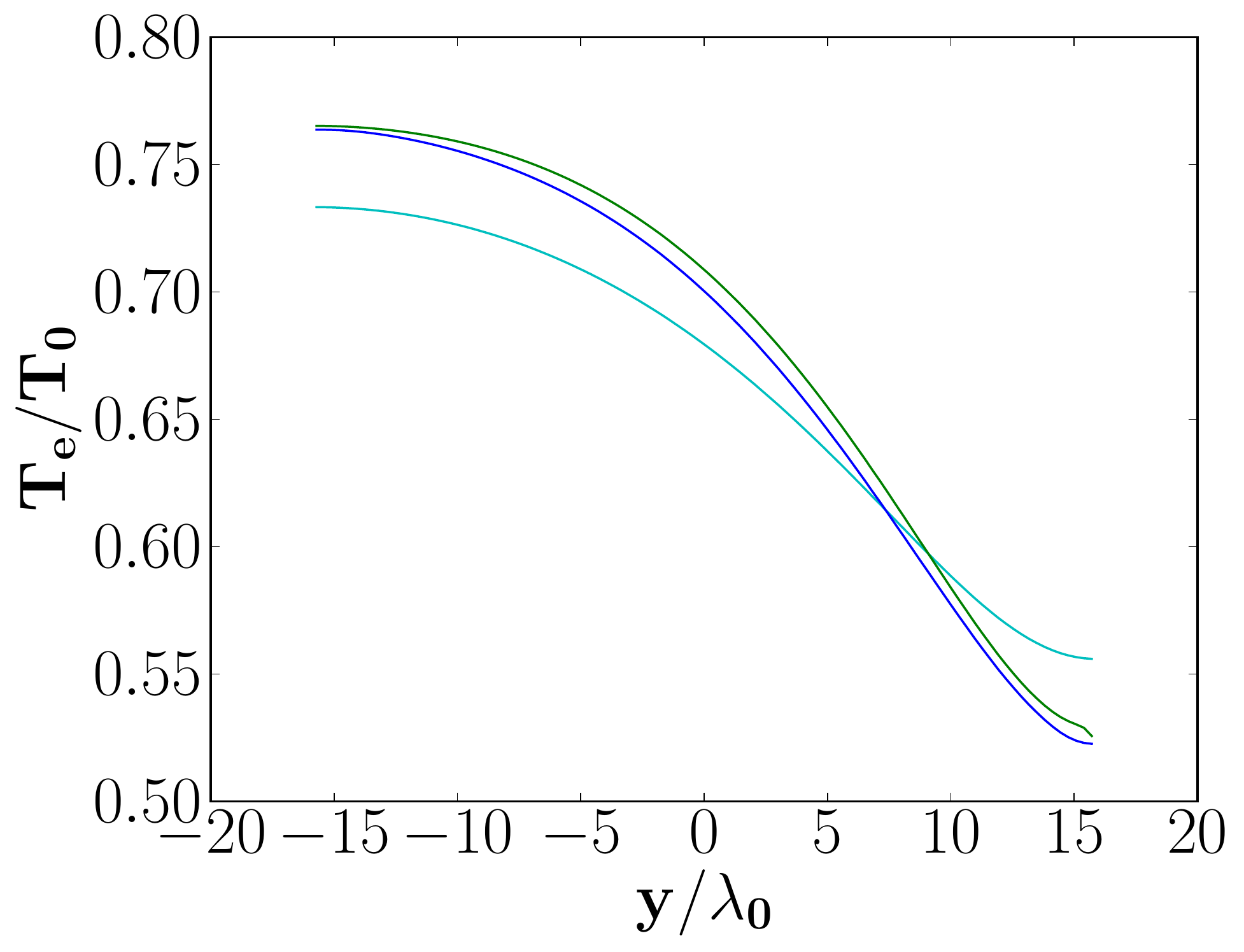}}~

    
    \caption{Cuts along the $x$ (a,c,e) and $y$-directions (b,d,f) of the temperature profiles shown in Fig.~\ref{2d_m1_B_ot=}, with $\omega_{B}\tau_{e}=0$ (a) and (b), $\omega_{B}\tau_{e}=0.1$ (c) and (d).  The case  $\omega_{B}\tau_{e}=0.5$  has not been shown since no sensible variations between models appear yet.  The temperature is predicted by the local (cyan lines), the SNB (green lines) and the M1 (blue lines) models.}
\end{figure*}

Dashed lines in Fig.~\ref{2d_m1_B_ot=} indicate the regions where line-outs have been performed. Cuts for an unmagnetized plasma along the $x$-direction are shown in Fig.~\ref{cut_y.eps}, while that along the $y$-direction in Fig.~\ref{cut_x.eps}. They show a general gradient smoothing, while a weak preheating is visible in front of the temperature gradient, along the $x$-direction, induced by nonlocal transport. The SNB model, also shown in the plot, agrees with M1 predictions.

Adding a weak degree of magnetization ($\omega_{B}\tau_{e}=0.1$) to the system, we observe the effect of flux rotation ($\sim5-10^{\circ}$, in the local case), related to magnetic fields. Temperature predictions are shown in Figs.~\ref{2d_brag_ot=01.pdf} (Braginskii's predictions), and \ref{2d_m1_B_ot=01.pdf} (M1 predictions). The SNB model provides results close to M1. Figures \ref{cut_y_ot=01.eps} and \ref{cut_x_ot=01.eps} show the horizontal and vertical cuts of the temperature profile. The Braginskii's theory predicts that the gradients are less smoothed than in the magnetized case. On the contrary, nonlocal models (M1, SNB) are not affected by such a weak magnetization. The nonlocal electrons behave as if they do not experience any magnetization. The SNB and M1 models give very similar results.

With $\omega_{B}\tau_{e}=0.5$, the effective MFP is reduced by a factor $\sim1/(1+C\omega_{B}^{2}\tau_{e}^{2})$, with $C\sim 1$, thus fluxes are reduced: the temperature gradients are less smoothed by the heat transport, as shown in Fig.~\ref{2d_brag_ot=05.pdf}. Despite a high degree of nonlocality, no differences can be found between Braginskii's, SNB and M1 predictions. The latter are shown in Fig.~\ref{2d_m1_B_ot=05.pdf}. 
 Because of the flux reduction, heating effects require a longer time to appear.

\begin{figure*}
    \centering
\subfloat[\label{tev_cut_y_ot=0.pdf}]{
        \centering
        \includegraphics[width=0.3\columnwidth]{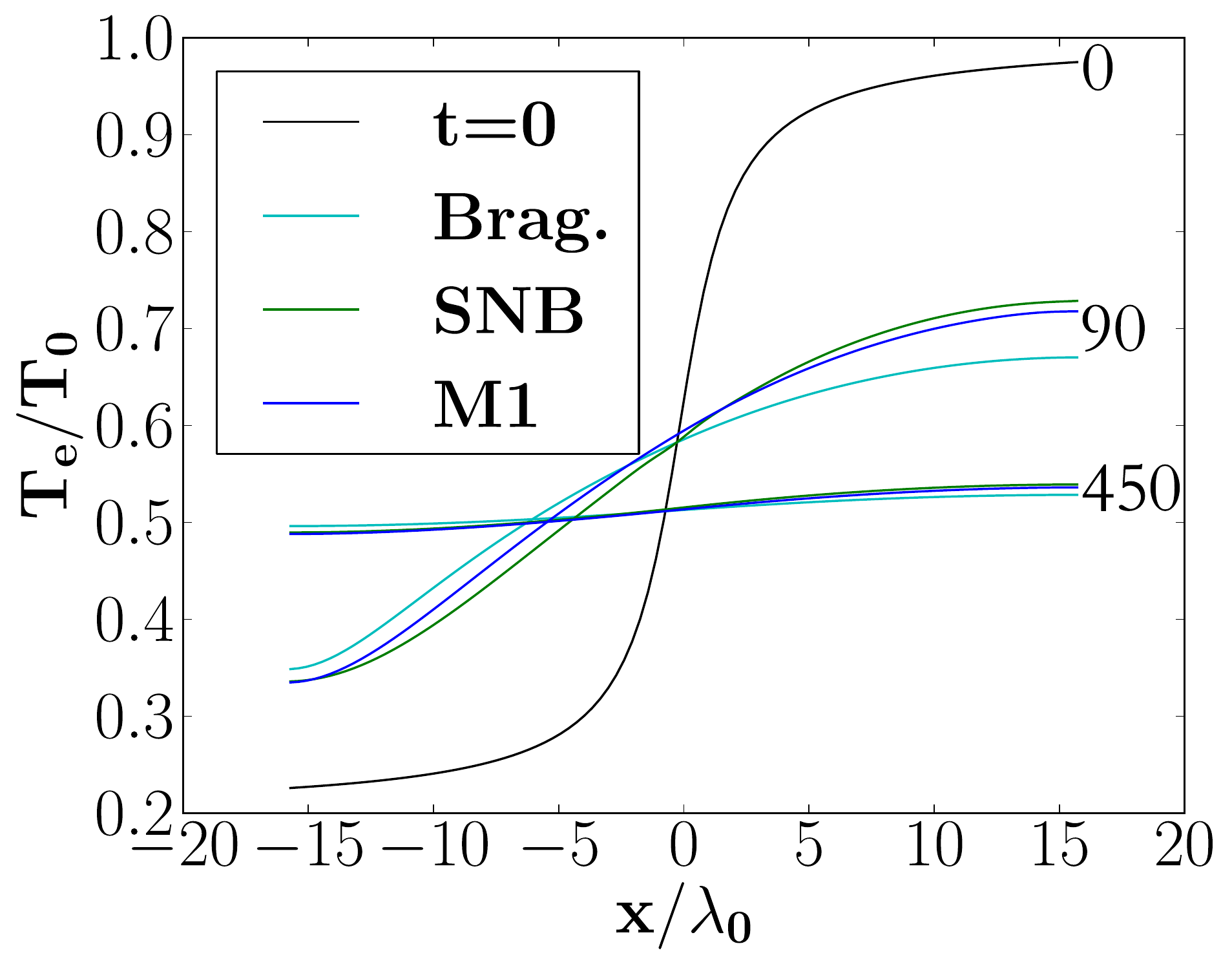}}~
\subfloat[\label{tev_cut_y_ot=01.pdf}]{
        \centering
        \includegraphics[width=0.3\columnwidth]{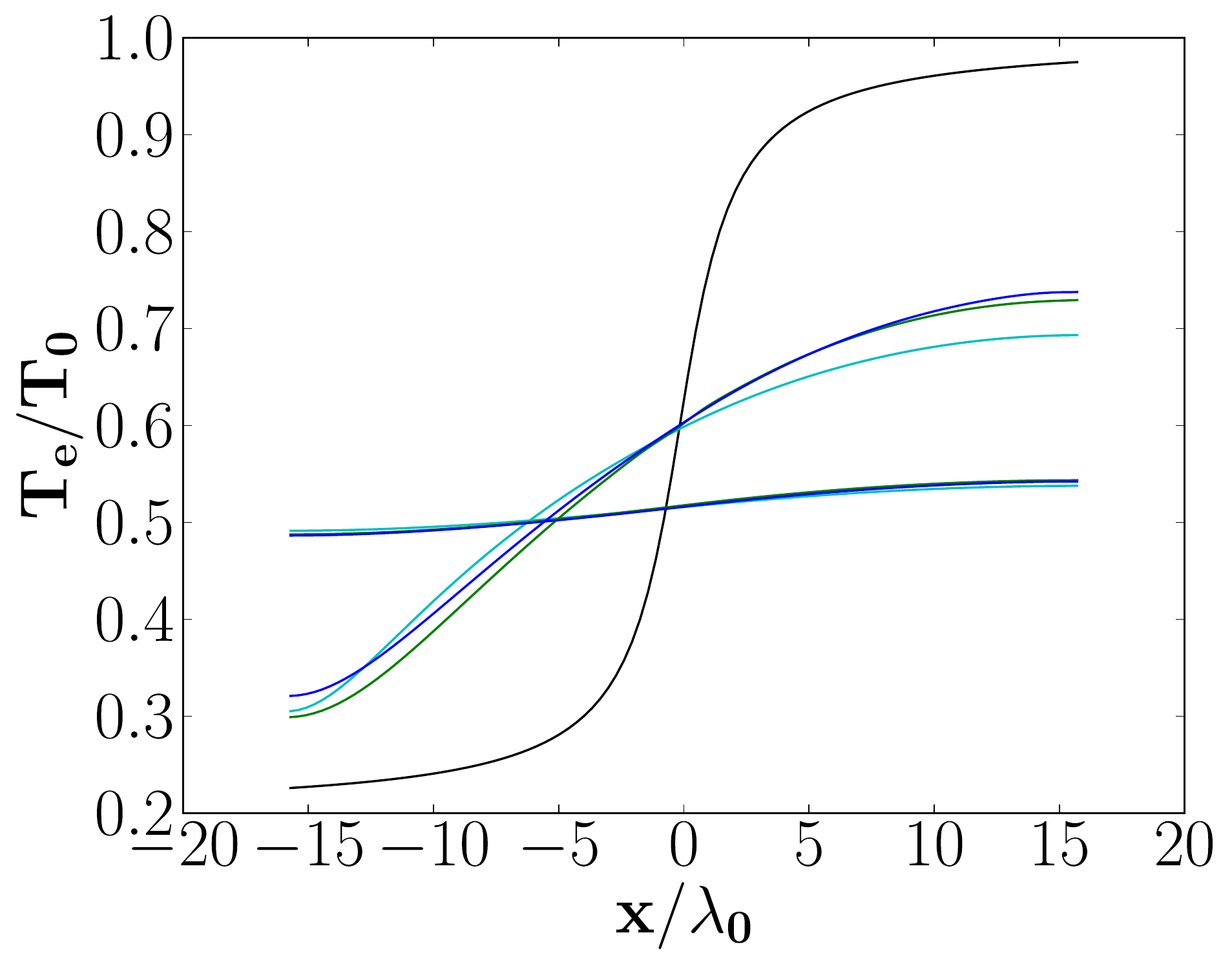}}~
\subfloat[\label{tev_cut_y_ot=05.pdf}]{
        \centering
        \includegraphics[width=0.3\columnwidth]{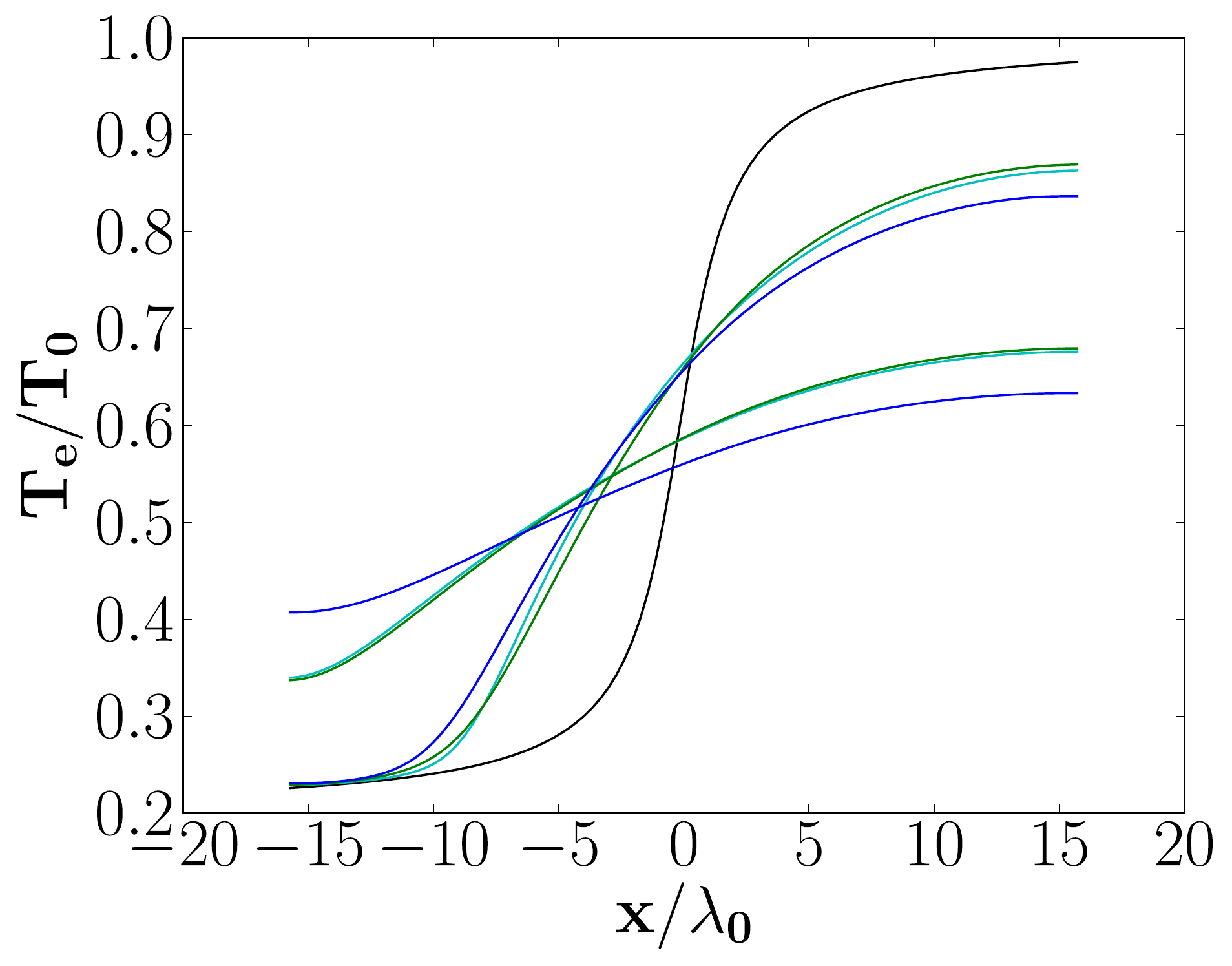}}

    \caption{Temporal evolution of the temperature profiles along the $x$-direction, for $\omega_{B}\tau_{e}=0$ (a), $0.1$ (b) and $0.5$ (c). The black curve represents the initial condition, the subsequent curves represent at $90\tau_{e}$ and $450\tau_{e}$, the temperature predicted by the local (cyan lines), the SNB  (green lines)  and the M1 (blue lines) models, for different degrees of magnetization. Black arrows have been added, in order to clarify the  variation of temperature in time.}
\end{figure*}

Figures \ref{tev_cut_y_ot=0.pdf}, \ref{tev_cut_y_ot=01.pdf} and \ref{tev_cut_y_ot=05.pdf} show a temporal evolution of the temperature profile, in the x-direction ($y\approx 14\lambda_{0}$), at 0, $90\tau_{e}$  (10 ps) and $450\tau_{e}$ ($50$ ps). As expected, temperature gradients decrease with time leading to a homogeneous temperature at the equilibrium. Considering the temperature difference evolution with time, $\Delta T_{e}(t)=\max[T_{e}(t)]-\min[T_{e}(t)]$, for $\omega_{B}\tau_{e}=0$, the local theory predicts a faster thermalization than the nonlocal model.  The SNB and M1 models  agree in this prediction. In the case of a weak magnetization $\omega_{B}\tau_{e}=0.1$, the thermalization time predicted by the Braginskii's theory is weakly reduced, while the Hall parameter is not sufficiently high to affect the nonlocal transport. So nonlocal models behave as in the case $\omega_{B}\tau_{e}=0$. 

A different result is obtained with a magnetization of $\omega_{B}\tau_{e}=0.5$. In this case, the magnetization is strong enough  to affect both local and nonlocal models. In the local simulations the thermalization time is strongly increased because the heat flux is reduced. The SNB model gives a local heat flux. On the contrary, the M1 model predicts a thermalization time longer than the local model. This result is surprising, since it is opposite to usual expectations from the nonlocal effects. Its explication is related to the heat flux rotation, shown in Fig.~\ref{t0_ot=05.pdf}. This rotation is weaker in the M1 simulation because of nonlocal effects, leading to a configuration in which the horizontal component of the flux in $x\approx0$ (main flux) is the same for the M1 model and for the Braginskii's theory. Nevertheless, in the nonlocal case, another flux is present: the preheating at $x\approx3\lambda_{0}$, which is absent in the local case. The preheating flux is responsible for the increase of heating efficiency in the horizontal direction.

Compared to the M1 model, the SNB model seems to overestimate the effects of magnetization. This difference
 becomes visible at long times, since magnetic effects reduce the heat transport, and so the temperature takes a longer time to be modified. It
is due to the different treatments of magnetic fields, which are phenomenological for the SNB model.  The system studied moves toward a lower degree of nonlocality, so differences are reduced in time. In the presence of a laser field, highly nonlocal regimes can be created. In this case, the differences, which now appear small, could become very important.

\section{Conclusions}\label{On the magnetized transport}

The M1 electron transport model is extended to the domain of magnetized plasmas.
It is able to reproduce results given by the Braginskii's theory, in the local regime. The nonlocal limit is studied for a wide range of cases. All of them show that magnetic field reduces nonlocal effects but also that nonlocal transport reduces magnetic effects. The analysis has been also performed at the  kinetic level, which demonstrates the flux rotation, in the phase space.

A thermal wave propagation through a magnetized plasma is performed. For weakly magnetized plasmas, the thermalization takes more time with the nonlocal model than with the local one because of the flux-limitation. On the contrary, in strongly magnetized plasmas, nonlocal effects are reduced.

The temporal analysis reveals a disagreement between the  M1 and the SNB model, which is explained by the fact that the latter overestimates the effects of magnetization, because of an approximate evaluation of the induced electric field effects.

In conclusion, the M1 model is able to deal with magnetized plasmas. 
It provides an efficient and robust method of description of suprathermal electron transport in high energy density plasmas,  remaining formally simpler than the generalization to magnetized plasmas of the SNB model.

\appendix

\section{Demonstration of thermal conduction dominance}\label{Demonstration of temperature dominance}

Here, we demonstrate the dominance of the thermal conduction, for the plasmas considered in Sec.~\ref{Local regime of magnetized plasmas}. This procedure can be extended to other plasmas in the limit of low magnetization.

The local heat flux in Eq.~\eqref{q braginskii} contains the thermal and thermoelectric components.
The thermoelectric heat flux is proportional to the current which has a nonzero value in the $y$-direction (crossed between the temperature gradient and the magnetic field). Since $\omega_{B}\tau_{e}$ is constant, $B_{z}\propto \tau_{e}^{-1} \propto T^{-3/2}_{e}$ and 
${\partial B_{z}}/{\partial x}=-{3}{B_{z}}/(2T_{e}){\partial T_{e}}/{\partial x}$. So, the current can be written as
\begin{equation}
j_{y}=\frac{3cB_{z}}{8\pi T_{e}}\frac{\partial}{\partial x}T_{e}
\end{equation}
and the $y$ component of the heat flux reads 
\begin{equation}
q_{By}=-k_{\wedge}\frac{\partial T_{e}}{\partial x}\left(1+\frac{\beta_{\perp}}{k_{\wedge}}\frac{3cB_{z}}{8\pi e}\right).
\end{equation}
The term in parenthesis can be expressed as a function of the Hall parameter and of the dimensionless thermal conductivity  $\bar{\bar{k}}^{c}=\bar{\bar{k}}m_{e}/(n_{e}T_{e}\tau_{e})$. Normalizing by the SH flux $q_{SH}(x)=-k^{c}_{\perp}(\omega_{B}\tau_{e}=0)n_{e}T_{e}\tau_{e}/m_{e}{\partial T_{e}}/{\partial x}$, we have 
\begin{equation}
\frac{q_{By}}{q_{SH}}=-\frac{k^{c}_{\wedge}}{k^{c}_{\perp}(\omega_{B}\tau_{e}=0)}\left(1+\frac{\beta_{\perp}}{k_{\wedge}^{c}}\omega_{B}\tau_{e}\frac{3}{2}\frac{m_{e}c^{2}}{T_{e}}\frac{\nu_{e}^{2}}{\omega_{pe}^{2}}\right).\label{formula trascuro beta}
\end{equation}

For the plasmas described in Sec.~\ref{Local regime of magnetized plasmas}, at the position $x=0$, $T_{e}=2.75\;\rm keV$,  $\omega_{pe}=1.8\times10^{16}\;\rm s^{-1}$ and $\tau_{e}=1.1\times10^{13}\;\rm s^{-1}$, for $Z=1$, and $\omega_{pe}=1.1\times10^{16}\;\rm s^{-1}$ and $\tau_{e}=2.9\times10^{14}\;\rm s^{-1}$, for $Z=79$.
Neglecting the thermoelectric contribution (second term in parenthesis), the error committed increases as $\omega_{B}\tau_{e}$ increases. In particular for the plasma with $Z=1$ and for $\omega_{B}\tau_{e}=30$, the error equals $0.06\%$. For the case  $Z=79$ and  for  $\omega_{B}\tau_{e}=3$, the error is around $5.4\%$ but for $\omega_{B}\tau_{e}>3$ the thermoelectric term needs to be accounted for.  
 Thus, in the limit of our analysis ($\omega_{B}\tau_{e}\leq30$ for $Z=1$ and $\omega_{B}\tau_{e}\leq3$ for $Z\gg1$), thermoelectric effects can be neglected and we only consider the thermal conductivity.

\section*{Acknowledgements}
This work has been carried out within the framework of the EUROfusion Consortium and has received funding from the European Unions Horizon 2020 research and innovation programme under Grant Agreement No. 633053.

    \bibliographystyle{plainnat}


 \end{document}